\documentclass[12pt]{article} 

\usepackage[T1]{fontenc}
\usepackage{kotex}
\usepackage{fvextra}

\usepackage{placeins}

\usepackage{threeparttable}
\usepackage{multirow}

\usepackage{listings}

\usepackage{libertine}

\usepackage{geometry} 
\usepackage{xcolor, graphicx} 

\usepackage{tikz}
\usetikzlibrary{positioning,arrows.meta}

\usepackage{multicol}

\usepackage{caption}

\usepackage{subcaption}

\usepackage{pgfplots}
\pgfplotsset{compat=1.16}

\usepackage{float}

\usepackage{caption}
\usepackage{tabularx}
\usepackage{booktabs}
\usepackage{float} 
\newcommand{\sym}[1]{\ensuremath{^{#1}}}
\usepackage{tabularx}
\usepackage{array} 

\usepackage{booktabs} 
\usepackage{array} 
\usepackage{paralist} 
\usepackage{verbatim} 
\usepackage{amsfonts, amsmath, amssymb, amsthm, euscript}
\usepackage{mathrsfs}
\usepackage{mhsetup, mathtools}
\usepackage{thmtools}
\usepackage{halloweenmath}

\theoremstyle{definition}

\renewcommand\thmcontinues[1]{Continued}

\definecolor{purple}{RGB}{85, 6,139}
\definecolor{teal}{RGB}{2,108,128}
\definecolor{lavender}{RGB}{129, 102, 122}
\definecolor{carolina blue}{RGB}{68, 157, 209}
\definecolor{phthalo blue}{RGB}{2, 8, 135}
\definecolor{purple2}{RGB}{149, 96, 219}
\definecolor{green1}{RGB}{96, 219, 117}
\definecolor{darkblue}{RGB}{0,0,102}

\usepackage{natbib}
\usepackage{hyperref}
\usepackage{nameref}
\hypersetup{
 breaklinks=true,
  colorlinks   = true, 
  urlcolor     = purple, 
  linkcolor    = darkblue, 
  citecolor   = darkblue, 
}

\usepackage{fancyhdr} 
\usepackage{sectsty}
\usepackage{setspace}
\allsectionsfont{\upshape \raggedright \fontsize{13}{15} \selectfont} 

\usepackage[nottoc,notlof,notlot]{tocbibind} 
\usepackage[titles,subfigure]{tocloft} 

\title{\textsc{Do People Follow AI Advice? \\
Evidence from a Pension Portfolio Choice Experiment}
\footnote{Hongseok Choi and Euncheol Shin were partly supported by Korea Institute for International Economic Policy. The authors are listed in alphabetical order and contributed equally to this paper.}
}
\author{
\begin{tabular}[t]{c@{\hskip 5em}c@{\hskip 5em}c}
 Hongseok Choi\footnote{Department of Economics, Sejong University, \href{mailto:aitchchoi@sejong.ac.kr}{aitchchoi@sejong.ac.kr}} &
 Jeongbin Kim\footnote{Department of Economics, Florida State University, \href{mailto:jkim33@fsu.edu}{jkim33@fsu.edu}} &
 Matthew Kovach\footnote{Department of Economics, Purdue University, \href{mailto:mlkovach@purdue.edu}{mlkovach@purdue.edu}} \\[0.5em]
 Kyu-Min Lee\footnote{Department of Management Information Systems, Chungbuk National University, \href{mailto:kyumin.lee@cbnu.ac.kr}{kyumin.lee@cbnu.ac.kr}} &
 Euncheol Shin\footnote{College of Business, Korea Advanced Institute of Science and Technology, \href{mailto:eshin.econ@kaist.ac.kr}{eshin.econ@kaist.ac.kr}} &
 Hector Tzavellas\footnote{Department of Economics, Virginia Tech, \href{mailto:hectortz@vt.edu}{hectortz@vt.edu}}
\end{tabular}
}
\date{
\today
}

\begin{document}

\maketitle

\vspace{5 mm}

\noindent{\textbf{Abstract:}
We study how differences in AI-generated financial recommendations are transmitted into individual portfolio choices. In an experiment with 400 employed adults enrolled in workplace defined contribution pension plans in South Korea, participants allocate a hypothetical pension balance across eleven products and may revise it after receiving one of two fixed AI-generated recommendations. A $2 \times 2$ design randomizes recommendation content and whether the recommendation includes a short rationale. Approximately 37\% of the experimentally induced difference between the aggressive and conservative recommendations passes through to final portfolios. This causal contrast changes expected portfolio return, volatility, allocations across risk grades, and the number of products held, but produces no detectable difference in computed Sharpe ratios. 81\% of participants revise. Among revisers, 95\% move toward the assigned recommendation and implement about half of the suggested adjustment. Rationales do not detectably alter pass-through. These results show that users partially and selectively transmit recommendation content into economically meaningful differences in risk exposure while retaining substantial weight on their initial choices.

}

\vspace{5 mm}


\vspace{3 mm}



\onehalfspacing

\pagebreak



\section{Introduction}
\label{section:introduction}

Artificial intelligence (henceforth, AI) can now generate advice in domains that involve consequential economic trade-offs, including consumption, saving, investment, and retirement planning \citep[e.g.,][]{NiszczotaandAbbas2023,LoandRoss2024}. As these systems become more accessible, individuals increasingly turn to them for guidance when making such decisions. Whether AI alters economic behavior, however, depends not only on the quality of the recommendations it produces, but also on whether and to what extent individuals actually follow them. Even an accurate recommendation affects outcomes only if users incorporate it into their decisions, while an inaccurate recommendation matters only to the extent that users act on it.

Much of the recent research on advice generated by general-purpose AI systems has focused on system capabilities, including whether these systems exhibit economic rationality, demonstrate financial literacy, or infer users' preferences from observed choices \citep[e.g.,][]{Chenetal:2023:WP,NiszczotaandAbbas2023,KimKovachLeeShinTzavellas2026}. This literature has substantially advanced our understanding of what AI systems can recommend and when their recommendations are more or less reliable. Much less is known, however, about the user side of the interaction: whether and how differences in AI-generated recommendations are transmitted into users' final decisions. This question is particularly important when decisions are multidimensional and lack a uniquely correct answer.\footnote{For recent work on human--algorithm interaction across management settings, see the \textit{Management Science} special issue introduced by \citet{caro2026humanalgorithm}.}

This response is not obvious. General-purpose AI occupies an unusual position as an advisor. It can process large amounts of information and provide rapid and confident guidance, but it lacks many of the characteristics associated with human advisors, such as credentials, experience, and accountability. Individuals may therefore defer to an AI recommendation, discount or ignore it, or combine it with their own prior judgment. Their response may also depend on how they assess their own judgment relative to that of the AI. How much of the content of an AI recommendation, then, passes through to individual decisions, and what explains variation in how people implement that advice?

We study these questions in the context of pension allocation. This setting is well suited to examining responses to AI-generated advice for several reasons. First, pension decisions are economically important for individual welfare because they shape long-run retirement outcomes and require difficult risk--return trade-offs \citep{campbell2006household}. Second, the task closely parallels a choice already faced by the population we study which consists of employed adults participating in workplace defined contribution (henceforth, DC) plans. Third, allocating a fixed balance across multiple products allows us to observe not only whether participants revise their decisions, but also whether their revisions point toward the recommendation and how much of the suggested adjustment they implement. Our objective is therefore not to study pension systems per se, but to use pension allocation as a consequential economic setting in which to examine how recommendation content is transmitted into individual choices.

We conduct an experiment with 400 employed adults in South Korea who are covered by DC workplace pension plans. Participants first allocate a hypothetical pension balance across eleven investment products using standardized information on past returns, risk grades, and return volatility. They are then shown an AI-generated portfolio recommendation and allowed to revise their initial allocation. The task includes a performance-based bonus, making both expected return and risk payoff-relevant.

The experiment uses a $2\times2$ between-subjects design. We randomly assign participants to receive one of two recommendations generated from identical product information. One is a relatively aggressive and concentrated portfolio produced by GPT-4 Turbo, while the other is a relatively conservative and diversified portfolio produced by GPT-4o.\footnote{At the time of the experiment in spring 2024, GPT-4 Turbo and GPT-4o were both frontier models. Although GPT-4o was newer, neither model uniformly dominated the other across tasks, and each had distinct performance characteristics \citep{OpenAI:2024:GPT4o,OpenAI:2024:GPT4oSystemCard}.} Participants are informed that the recommendation was generated by an AI system but are not told which model produced it. We therefore interpret this treatment as variation in the substantive content of the advice rather than as a comparison of model identities. Independently, we vary whether the numerical recommendation is presented alone or accompanied by a short rationale explaining the logic of the suggested portfolio.\footnote{The rationale manipulation is motivated by evidence that explanations can reshape how users respond to algorithmic output \citep{bauer2023explained}.}

Because participants are randomly assigned to one of two substantively different recommendations, the experiment allows us to measure how much of the difference in advice is reflected in their final choices. We place each portfolio on the line connecting the conservative and aggressive recommendations, normalizing the conservative recommendation to zero and the aggressive recommendation to one. The randomized difference in average final positions then measures the share of the experimentally induced difference in recommendation content that passes through to participants' portfolios. This approach identifies the effect of assignment to these two specific recommendation portfolios; it does not identify a general effect of one GPT model relative to another.

Our analysis yields three main findings.

\begin{itemize}

\item \textbf{Substantial but incomplete causal pass-through.}
Approximately 37 percent of the difference between the aggressive and conservative recommendations passes through to participants' final portfolios. The estimate is highly stable across change-score, baseline-adjusted, and covariate-adjusted specifications, and we reject both no pass-through and full pass-through. Recommendation content therefore has a substantial causal effect on portfolio choice, but participants transmit only part of the experimentally induced difference in advice. We find no statistically reliable evidence that the short rationale either shifts average portfolio positions or changes the degree of pass-through.

\item \textbf{Pass-through changes risk exposure rather than risk-adjusted performance.}
The causal effects closely reflect the economic content of the recommendations. Assignment to the aggressive recommendation raises expected return and portfolio risk, reduces allocations to principal-protected and low-risk products, increases allocations to high- and very-high-risk products, and reduces the number of products held. Across these dimensions, approximately 28 to 40 percent of the difference between the recommendations appears in participants' final portfolios. By contrast, differences in the recommendations' Sharpe ratios and concentration measures do not pass through detectably. AI advice therefore changes the level and composition of risk that participants assume, without producing a detectable improvement in risk-adjusted performance.

\item \textbf{Incomplete pass-through reflects distinct margins of individual implementation.}
Eighty-one percent of participants revise their portfolios after receiving the recommendation. Among those who revise, 95 percent move in a direction aligned with the recommendation and implement approximately half of the suggested adjustment. Incomplete aggregate pass-through therefore reflects both an extensive margin---nearly one fifth of participants do not revise---and an intensive margin---active revisers move only partway and also make changes orthogonal to the advice. Investor characteristics are associated with different components of this process. Participants who mistakenly believe that their baseline portfolio dominates the recommendation revise by as much as others but direct less of their revision toward the AI. By contrast, participants whose baseline portfolio is less efficient than the recommendation revise more, but do not end significantly closer to it. Because these characteristics are not randomly assigned, and beliefs are elicited after the decision, these relationships are descriptive rather than causal.

\end{itemize}

Taken together, the findings show that the influence of AI advice is not well described as a binary choice between acceptance and rejection. At the aggregate level, differences in recommendation content pass through substantially but incompletely to final choices. At the individual level, most participants respond in the recommended direction, but they retain considerable weight on their initial judgments and frequently make additional adjustments not prescribed by the AI. The multidimensional structure of the task allows us to connect these two levels of analysis through a causal measure of aggregate pass-through and an individual-level decomposition of whether participants revise, where they move, and how far they go.

These results have implications for the deployment of AI advisory systems. The effects of AI advice depend jointly on what a system recommends and on how users translate the recommendation into action. Two general-purpose models given identical information can produce materially different advice, and a substantial portion of those differences can be transmitted to users' decisions even when neither recommendation is followed fully. Model and system design are therefore consequential not only because they affect the quality of recommendations, but also because their substantive content can shape the risks users ultimately bear. More broadly, general-purpose AI may influence economic decisions not by replacing human judgment, but by providing a direction that individuals follow partially and selectively.


\vskip+1em

\noindent \textbf{Related literature.} Our paper contributes first to the literature on how individuals use algorithmic advice. Existing evidence is mixed. Some studies document \emph{algorithm aversion}, whereby individuals rely less on algorithmic than on human judgment even when the algorithm performs better \citep{dietvorst2015algorithm,castelo2019algorithm}. Others find \emph{algorithm appreciation} or show that resistance declines when users retain some control over the final decision \citep{dietvorst2018algorithm,logg2019algorithm}.\footnote{See \citet{Burton2020AlgorithmAversion} and \citet{caro2026humanalgorithm} for broader discussions of human--algorithm interaction.} Because all participants in our experiment receive AI advice, and we do not compare it with advice from a human, our design is not a direct test of algorithm aversion. Instead, we study how individuals integrate a nonbinding AI recommendation with their own prior judgment once the advice is available.

In this respect, our analysis connects closely to the advice-taking literature. When individuals receive advice from another person, they typically move toward it but place excessive weight on their own initial judgment \citep{harvey1997taking,yaniv2000advice}. More recently, \citet{Balakrishnan2026NaiveAdvice} document a similar pattern in human--algorithm collaboration. They show that decision makers combine their own forecast with an algorithmic prediction using a weight that responds only weakly to the relative informativeness of the two signals. 

We extend this literature in two ways. First, random assignment to two substantively different recommendation portfolios allows us to estimate the causal pass-through of recommendation content, i.e., the fraction of the experimentally induced difference between the recommendations that appears in participants' final choices. Second, our setting involves a multidimensional, jointly constrained allocation rather than a scalar estimate. We therefore complement the aggregate pass-through measure with an individual-level decomposition that separates whether investors revise, whether their revisions point toward the recommendation, and how much of the recommended adjustment they implement. The two approaches connect variation in the content of advice to the behavioral process through which individuals incorporate it.

Our paper also contributes to the emerging literature on AI-generated financial advice. The paper most closely related to ours is \citet{yang2026advisor}, who conduct a field experiment in which customers of a savings bank receive AI-based advice before making a series of investment decisions. Their system is task-specific, whereas we study recommendations produced by general-purpose large language models that retail investors can consult directly. Their participants make a sequence of binary investment decisions, while ours allocate a fixed balance across multiple assets. Moreover, because our experiment randomly assigns participants to two recommendations that differ substantially in content, we can estimate how much of the difference between those recommendations passes through to final choices. The jointly constrained portfolio task also allows us to trace this pass-through across expected return, risk exposure, portfolio composition, breadth, concentration, and risk-adjusted performance.\footnote{See also \citet{NiszczotaandAbbas2023}, who provide preliminary evidence on responsiveness to GPT advice in a simple savings problem, and \citet{oehler2024gpt}, who compare portfolio recommendations generated by ChatGPT and robo-advisors but do not study whether investors act on those recommendations.}

The paper is further related to research on robo-advising. \citet{bianchi2024humanrobot} study the introduction of robo-advice in employee saving plans and find that it increases investor attention and trading. Other studies examine services that construct or manage portfolios on the investor's behalf \citep{reher2024robo,rossi2024diversification}. In particular, \citet{rossi2024diversification} find that robo-advice improves diversification and risk-adjusted performance, with larger gains among investors whose initial portfolios are less diversified. These systems are specialized financial products and often involve an ongoing advisory or delegated-management relationship. By contrast, we study how individuals respond to a one-time, nonbinding recommendation generated by a general-purpose AI. The comparison is informative because participants in our experiment respond substantially to the advice, yet the resulting changes primarily alter risk exposure rather than producing robust improvements in risk-adjusted performance.

Finally, our setting connects the paper to the literature on household portfolio choice and financial advice. Households make consequential investment mistakes, including underdiversification and nonparticipation in risky asset markets \citep{campbell2006household,calvet2007down}. Retirement decisions are also characterized by default-driven inertia \citep{madrian2001power} and naive menu-splitting \citep{benartzi2001naive}. At the same time, investors who may benefit most from financial guidance are often the least likely to obtain or follow it \citep{bhattacharya2012advice}, while human financial advisors provide limited tailoring of portfolio allocations to client characteristics \citep{foerster2017advice}. These patterns make retirement portfolio allocation a natural setting in which to study whether accessible AI-generated advice changes behavior and whether those changes improve financial decisions.

\section{Experimental Design}
\label{section:design}

\subsection{Participants and Eligibility}

We recruit 400 adult participants in South Korea between the ages of 35 and 55 who are currently employed and covered by a workplace DC pension plan. Screening questions exclude those who are not employed, whose workplace offers no retirement plan, or who are covered only by defined benefit or individual retirement pension accounts.\footnote{Eligibility was further restricted by occupation: only company employees, civil servants (including public enterprises), teachers and instructors, licensed professionals (e.g., professors, lawyers, physicians), and those in agriculture, forestry, or fishing continued. Self-employed persons, freelancers, homemakers, students, and the unemployed were screened out. Participants were recruited by Macromill Embrain, a professional survey company based in South Korea that has been used in economics surveys and experiments \citep[e.g.,][]{chao2026culture, JungandKim2022}. Embrain maintains a standing online participant pool and can recruit participants while controlling the sample composition, for example, by age and device type.} We focus on DC participants because they bear the investment risk of their own portfolios and actively choose allocations, so the experimental task parallels a decision they already face. The 35--55 band captures employees with meaningful accumulated balances who remain far enough from retirement that risk-return trade-offs are economically relevant. Sampling quotas of 25 respondents in each gender (male, female) by age group (35--45, 46--55) cell for every treatment group yield 100 per treatment and 400 in total, balancing demographics across treatments. Participation is restricted to desktop or laptop devices because the eleven-product table cannot be displayed legibly on mobile screens.


\subsection{Treatments}

The experiment uses a between-subjects $2 \times 2$ design, as illustrated in \autoref{tab:treatments}. All participants complete the same two-stage pension allocation task. In the first stage, they choose an initial portfolio based only on information about the eleven pension products used in \citet{HaKimKimShin2019}.\footnote{We use the 11-product menu from \citet{HaKimKimShin2019}, which is designed to reflect the risk-return profiles of actual Korean retirement pension products. The menu therefore provides a realistic but controlled choice environment for studying retirement portfolio decisions among Korean subjects. The details of the products are presented in \autoref{tab:investment_products}.} In the second stage, they see an AI-generated recommendation and are allowed to revise their portfolio. The treatment conditions vary along two dimensions: which GPT model generates the recommendation and whether the recommendation is accompanied by a verbal rationale.

\begin{table}[!ht]
\centering
\footnotesize
\caption{Treatment Conditions (2$\times$2 Design)}
\label{tab:treatments}
\begin{tabular}{llcc}
\toprule
    &                     & \multicolumn{2}{c}{Rationale} \\
\cmidrule(lr){3-4}
    &                     & No              & Yes             \\
\midrule
\multirow{2}{*}{GPT version} 
    & GPT-4 Turbo          & Group 1         & Group 2         \\
    & GPT-4o               & Group 3         & Group 4         \\
\bottomrule
\end{tabular}
\end{table}

The first dimension is the version of GPT used to generate the recommendation portfolio. In two conditions, the recommendation is generated by GPT-4 Turbo, and in the other two conditions it is generated by GPT-4o.\footnote{These were two state-of-the-art versions of OpenAI's GPT models. Participants were not informed of the model version; they were only told that the recommendation was generated by an AI system. \autoref{appendix:prompt} provides the prompts used to generate the recommendations and rationales in detail. We wrote the prompts in Korean to avoid potential language or cultural bias that could arise from using English, which was not the subjects' first language.} In each case, the model returns a complete allocation vector over the eleven products based on the same product information table shown to participants. As explained in the following subsection, the GPT-4 Turbo recommendation places relatively high weight on products with high historical returns and higher risk, especially Products 8 and 11, while the GPT-4o recommendation assigns relatively more weight to lower risk products and less weight to the riskiest products.\footnote{Each model's recommendation was generated by querying the model 100 times at temperature 0.5 with the identical prompt in \autoref{appendix:prompt}. We used the modal allocation across the 100 draws, $(10,10,5,15,0,0,10,20,0,0,30)$ for GPT-4 Turbo (32 of 100 draws) and $(20,20,10,10,10,10,5,5,0,5,5)$ for GPT-4o (16 of 100 draws). In the rationale conditions, the rationale was drawn at random from the generations producing the modal allocation. The lower modal frequency for GPT-4o reflects dispersion across identical queries, consistent with evidence of lower output consistency even at temperature zero \citep[e.g.,][]{klishevich2025determinism}, though we do not test determinism directly.} Since the two models produce systematically different portfolios from identical information, we treat the model dimension as a manipulation of advice content, aggressive versus conservative, rather than a comparison of models per se.

The second dimension is whether a short rationale is shown together with the numerical recommendation. In the recommendation-only conditions, participants see only the recommended weights for each product. In the recommendation-with-rationale conditions, the same numerical recommendation is displayed together with a model-specific paragraph that explains, in qualitative terms, how that model's recommended portfolio balances higher-return and more stable products, and why some products receive low or zero weight. This additional dimension is designed to capture an important feature of real-world interactions with \textit{generative} AI: users often receive not only numerical recommendations but also verbal rationales for those recommendations. Such rationales may change how participants interpret and evaluate the recommendation. This treatment variation therefore allows us to separate the effect of the numerical advice from the additional effect of its verbal explanation. Crossing these two dimensions yields four treatment groups, summarized in \autoref{tab:treatments}. The model treatment changes the content of the advice: GPT-4 Turbo recommends more aggressive portfolios, whereas GPT-4o recommends more conservative ones. The rationale treatment changes only how the advice is presented. Analysis plans were preregistered on AsPredicted prior to data collection.\footnote{The preregistration is publicly available at \url{https://aspredicted.org/7km5-4yxx.pdf}.}


\subsection{Task and Information}
\label{subsection:taskninfo}

The core task is a hypothetical but incentivized allocation of a DC pension balance across eleven products. As shown in \autoref{tab:investment_products}, each product is described in a table with the following columns: multi-horizon gross returns (three-month, six-month, one-year, two-year, and three-year horizons), a discrete risk grade, the product-level standard deviation of returns, and an input field for the investment share as a percentage of the pension balance.\footnote{The actual screenshot of this table, together with additional background details, is provided in \autoref{appendix:additional_figures}.} Participants are instructed to think of the task as a realistic decision about their own retirement income and to choose the allocation that they find appropriate for their retirement planning. The investment shares must sum to one hundred, and participants may leave any product at zero if they prefer not to invest in it. Client-side and server-side checks enforce the one hundred percent constraint.

\begin{table}[htbp]
\centering
\footnotesize
\caption{Investment Products}
\label{tab:investment_products}
\begin{tabular}{ccccccccr}
\hline
\multirow{2}{*}{Product} 
& \multicolumn{5}{c}{Return (\%)} 
& \multirow{2}{*}{Risk grade} 
& \multirow{2}{*}{Std. dev.} 
& \multirow{2}{*}{Weight} \\
\cline{2-6}
& 3 months & 6 months & 1 year & 2 years & 3 years 
& & & \\
\hline
Product 1  &  0.19 &  0.40 &  0.81 &  1.65 &  2.51 & 5 (Low risk) &  0.09 & 0\% \\
Product 2  & -0.02 &  0.32 &  0.74 &  2.00 &  2.84 & 5 (Low risk) &  1.09 & 0\% \\
Product 3  & -0.20 & -0.16 &  2.14 &  4.07 &  4.13 & 4 (Moderate risk) &  6.30 & 0\% \\
Product 4  &  0.81 &  3.17 &  4.86 & 12.27 & 19.27 & 4 (Moderate risk) &  9.65 & 0\% \\
Product 5  & -0.66 &  0.22 & -1.00 &  5.55 &  9.69 & 3 (Moderately high risk) & 10.91 & 0\% \\
Product 6  & -3.23 & -0.16 &  5.88 & 10.12 &  7.88 & 3 (Moderately high risk) & 14.43 & 0\% \\
Product 7  &  3.23 &  3.57 & 17.95 & 17.76 & 13.65 & 2 (High risk) & 24.24 & 0\% \\
Product 8  &  8.36 &  9.52 & 16.67 & 25.53 & 33.27 & 2 (High risk) & 21.01 & 0\% \\
Product 9  & -3.73 & -3.84 & -8.10 & -3.79 &  0.90 & 2 (High risk) & 21.31 & 0\% \\
Product 10 & -5.41 & -4.23 &  5.50 &  4.28 &  4.43 & 1 (Very high risk) & 28.24 & 0\% \\
Product 11 &  5.73 &  3.69 & 16.75 & 22.39 & 25.72 & 1 (Very high risk) & 27.09 & 0\% \\
\hline
Total & & & & & & & & 0\% \\
\hline
\end{tabular}
\vspace{2pt}
\begin{minipage}{\textwidth}
\footnotesize
\textit{Notes:} A larger standard deviation indicates greater return volatility.
The expected annual inflation rate is 1.2 percent, and the annual retirement pension management fee is 0.6 percent.
Past returns on performance-based products do not guarantee future returns.
\end{minipage}
\end{table}

Before making any allocation, participants see an introductory screen that explains the basic features of defined benefit and DC pension plans in Korea and clarifies that the task concerns a DC plan. They then see the product table together with short explanations of returns, risk grades, and standard deviation.


\subsection{Procedure and Survey Flow}

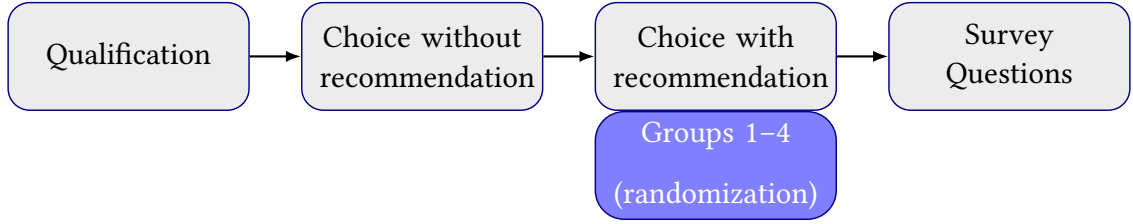
\begin{figure}[ht]
\centering
\resizebox{0.90\textwidth}{!}{%
\begin{tikzpicture}[
    >=Latex,
    thick,
    node distance=1.2cm and 1.0cm,
    stage/.style={
        draw=blue!50!black,
        rounded corners=12pt,
        fill=gray!15,
        minimum width=4.7cm,
        minimum height=2.1cm,
        text width=4.0cm,
        align=center,
        font=\Large
    },
    randombox/.style={
        draw=blue!50!black,
        rounded corners=16pt,
        fill=blue!50,
        minimum width=4.7cm,
        minimum height=2.1cm,
        text width=4.0cm,
        align=center,
        font=\Large,
        text=white
    }
]

\node[stage] (qual) {Qualification};
\node[stage, right=of qual] (choice0) {Choice without \\recommendation};

\node[stage, right=of choice0] (choice1) {Choice with \\recommendation};
\node[randombox, below=0cm of choice1.south, anchor=north] (groups) {Groups 1--4\\[0.4cm](randomization)};

\node[stage, right=of choice1] (survey) {Survey\\Questions};

\draw[->, line width=1.2pt] (qual.east) -- (choice0.west);
\draw[->, line width=1.2pt] (choice0.east) -- (choice1.west);
\draw[->, line width=1.2pt] (choice1.east) -- (survey.west);

\end{tikzpicture}%
}
\caption{Experimental procedure}
\label{fig:procedure}
\end{figure}

The survey proceeds in four steps, as shown in \autoref{fig:procedure}, and takes about 20 minutes to complete.\footnote{The study was approved by the IRBs of KAIST and Florida State University. All participants gave informed consent.} First, participants read an informed consent statement that explains the purpose of the study, the voluntary nature of participation, the use of anonymous data for academic research, and the payment scheme. Only those who agree continue to the main task.

Second, participants enter the baseline decision stage. They are shown the pension product information screen and answer three short comprehension questions that test their understanding of negative returns and capital loss, relative risk across products based on standard deviation, and which product has the highest expected three-year return. These questions are used to measure basic understanding of the information but do not affect payment. Then, participants choose an initial allocation vector across the eleven products (without recommendation). They enter a percentage for each product subject to the one hundred percent constraint. Each weight is entered as an integer in [0,100]. The platform rejects decimals and out-of-range values and blocks progress until the eleven weights sum to exactly 100. The resulting baseline allocation is stored as $w = (w_1,\dots,w_{11})$.

Third, participants proceed to the choice with recommendation stage. At this stage, the survey platform randomly assigns participants to one of the four treatment conditions, subject to the age-by-gender quotas described above. They are informed that an AI system has constructed a recommended allocation for each product based on the same information table. The same product table is shown again. Additionally, as shown in \autoref{tab:investment_products_recommendation}, one column displays the participant's own baseline allocation, the next column displays the AI recommendation corresponding to the participant's treatment condition, and the last column allows participants to choose a new allocation. The AI-recommended allocation varies across treatment conditions. In the recommendation-with-rationale conditions (Groups 2 and 4), a short paragraph below the table explains the logic of the recommendation in terms of diversification and the trade-off between stability and return.\footnote{See \autoref{appendix:additional_figures} for a screenshot of the product menu; the recommendation stage displayed the same table with two additional columns showing the participant's baseline allocation and the AI-recommended weights.} Participants are again required to choose investment shares that sum to one hundred. The resulting final allocation is stored as $v = (v_1,\dots,v_{11})$, together with the corresponding recommendation vector $r = (r_1,\dots,r_{11})$. The recommendation vector by GPT-4 Turbo is $(10,10,5,15,0,0,10,20,0,0,30)$, and the vector by GPT-4o is $(20,20,10,10,10,10,5,5,0,5,5)$. Random assignment produced well-balanced cells. Baseline portfolio moments and participant characteristics do not differ significantly across the four conditions, as reported in the balance table in Appendix \ref{appendix:balance}.

\begin{table}[!ht]
\centering
\footnotesize
\caption{Investment Products with AI Recommendation (GPT-4 Turbo)}
\label{tab:investment_products_recommendation}
\resizebox{\textwidth}{!}{%
\begin{tabular}{ccccccccccc}
\hline
\multirow{2}{*}{Product} 
& \multicolumn{5}{c}{Return (\%)} 
& \multirow{2}{*}{Risk grade} 
& \multirow{2}{*}{Std. dev.} 
& \multirow{2}{*}{\begin{tabular}{c}Previous\\weight\end{tabular}}
& \multirow{2}{*}{\begin{tabular}{c}AI \\weight\end{tabular}}
& \multirow{2}{*}{Weight} \\
\cline{2-6}
& 3 months & 6 months & 1 year & 2 years & 3 years 
& & & & & \\
\hline
Product 1  &  0.19 &  0.40 &  0.81 &  1.65 &  2.51 & 5 (Low risk) &  0.09 & 100\% & 10\% & 0\% \\
Product 2  & -0.02 &  0.32 &  0.74 &  2.00 &  2.84 & 5 (Low risk) &  1.09 & 0\% & 10\% & 0\% \\
Product 3  & -0.20 & -0.16 &  2.14 &  4.07 &  4.13 & 4 (Moderate risk) &  6.30 & 0\% & 5\% & 0\% \\
Product 4  &  0.81 &  3.17 &  4.86 & 12.27 & 19.27 & 4 (Moderate risk) &  9.65 & 0\% & 15\% & 0\% \\
Product 5  & -0.66 &  0.22 & -1.00 &  5.55 &  9.69 & 3 (Moderately high risk) & 10.91 & 0\% & 0\% & 0\% \\
Product 6  & -3.23 & -0.16 &  5.88 & 10.12 &  7.88 & 3 (Moderately high risk) & 14.43 & 0\% & 0\% & 0\% \\
Product 7  &  3.23 &  3.57 & 17.95 & 17.76 & 13.65 & 2 (High risk) & 24.24 & 0\% & 10\% & 0\% \\
Product 8  &  8.36 &  9.52 & 16.67 & 25.53 & 33.27 & 2 (High risk) & 21.01 & 0\% & 20\% & 0\% \\
Product 9  & -3.73 & -3.84 & -8.10 & -3.79 &  0.90 & 2 (High risk) & 21.31 & 0\% & 0\% & 0\% \\
Product 10 & -5.41 & -4.23 &  5.50 &  4.28 &  4.43 & 1 (Very high risk) & 28.24 & 0\% & 0\% & 0\% \\
Product 11 &  5.73 &  3.69 & 16.75 & 22.39 & 25.72 & 1 (Very high risk) & 27.09 & 0\% & 30\% & 0\% \\
\hline
Total & & & & & & & & 100\% & 100\% & 0\% \\
\hline
\end{tabular}%
}
\vspace{2pt}
\begin{minipage}{\textwidth}
\footnotesize
\textit{Notes:} A larger standard deviation indicates greater return volatility. The expected annual inflation rate is 1.2 percent, and the annual retirement pension management fee is 0.6 percent. Past returns on performance-based products do not guarantee future returns.
\end{minipage}
\end{table}

Fourth, after completing the allocation task, participants answer a series of follow-up questions about which pieces of information they relied on most (returns at different horizons versus the recommendation), how they perceive the risk and return of their original and final allocations relative to the AI recommendation and relative to their own original allocation, and how similar their hypothetical choices are to their actual pension choices. Further questions elicit whether they know the return on their own pension, how well they understand the risk of their pension, whether they know how to change pension products, and whether they are enrolled in an individual retirement pension account. Finally, participants complete a battery of incentivized quiz questions measuring financial literacy (five items on compound interest, inflation, the time value of money, and money illusion, adapted from \citet{vanrooij2011financial}) and cognitive reflection (six-item CRT-Long of \citet{primi2016development}, extending \citet{frederick2005cognitive}), a set of questions on attitudes toward AI products and services, general risk tolerance and time preference on zero to ten scales, and demographic questions on education, income, household income, and household size.

The survey platform also implements several checks to ensure consistent data collection. Device checks prevent participation from mobile phones and tablets, because the eleven-product table cannot be displayed legibly on smaller screens. The one hundred percent allocation constraint is enforced both on the client side and on the server side, and participants cannot proceed until a valid allocation is entered. All allocation variables, timestamps, and responses to follow-up questions are stored in a consistent format for analysis. 


\subsection{Incentives, Outcomes, and Measures}

\noindent \textbf{Incentives.} All participants received the standard fixed participation payment from the survey company. The allocation task also included an additional performance-based bonus. Participants were informed that their chosen portfolio would be used in a computer simulation of retirement wealth. For each product, the simulation drew annual returns from a normal distribution calibrated to the product's return and standard deviation as reported in the menu, independently across products, and compounded the portfolio return over a participant-specific horizon, defined as the number of years remaining until age 60. The simulated retirement wealth determined an additional payment between 500 and 30,000 points, where one point equals one Korean won. Participants were also told that this bonus was separate from the fixed survey payment.\footnote{For example, the financial-literacy and cognitive-reflection quiz carried a separate incentive: one participant was drawn to receive an additional 20,000 points (20,000 KRW), with the winning probability increasing in the number of items answered correctly. This lottery was independent of the fixed payment and the allocation-task bonus.} Since returns were drawn with product-specific volatilities, the riskiness of the chosen portfolio translated directly into the variability of the bonus, so that both the expected return and the risk of the allocation were payoff-relevant. The bonus was paid by the survey company within a few business days after completion of the survey.

\vskip+1em

\noindent \textbf{Outcomes.} The primary data are the two allocation vectors each participant chooses. We observe the baseline portfolio $w$, chosen before the recommendation, the final portfolio $v$, chosen after it, and the treatment-specific recommendation vector $r$. Weights are expressed on a $[0,1]$ scale. Our first outcome is proximity to the advice. The \emph{baseline distance} $d^{0}=\lVert w-r\rVert$ and the \emph{final distance} $d^{1}=\lVert v-r\rVert$ are the Euclidean distances between each portfolio and the received recommendation. The gap closed, $d^{0}-d^{1}$, measures \emph{adherence}; that is, how far the participant moves toward the AI. Some tables report the change in distance $d^{1}-d^{0}$, the negative of the gap closed. Our second outcome is the \emph{revision magnitude} $\lVert v-w\rVert$, which captures how much the participant changes the portfolio regardless of direction. We further decompose each revision below into a component directed at the recommendation and an orthogonal component. Our third set of outcomes describes portfolio quality: the expected return, standard deviation, and Sharpe ratio of $w$ and $v$, and their changes.\footnote{Each product's expected annual return is the geometric annualization of its three-year menu return, $[(1+r^{3\mathrm{y}}/100)^{1/3}-1]\times 100$. Portfolio risk uses the menu standard deviations and treats product returns as uncorrelated, because the menu reports no correlations. The Sharpe ratio is $(\mu_p-\mu_1)/\sigma_p$, where $\mu_1$ is the annualized return of Product~1, the only principal-guaranteed product, which serves as our proxy for the risk-free rate. Product 1 is not treated as riskless: its standard deviation of $0.09$ percent enters portfolio risk like that of any other product. Figures report returns in real terms, net of $1.8$ percentage points per year ($1.2$ for expected inflation and $0.6$ for the management fee). The zero-correlation assumption affects the portfolio standard deviation, the Sharpe ratio, the objective dominance indicator, and the efficient frontier, but not our main outcome measures: distance, revision magnitude, directedness, and the weight on advice depend only on portfolio weights and are therefore invariant to the correlation structure. Imposing a common pairwise correlation of 0.2 removes the small average Sharpe gain reported below, reverses the Sharpe ranking of the two recommendations, and raises the share of participants whose baseline portfolio objectively dominates the recommendation from 3 to 6 percent, leaving the contrast with the 27 percent who believe it does.}

\vskip+1em

\noindent \textbf{Measures.} The post-task survey collects the covariates used in the analysis. Its key item elicits beliefs about the recommendation. Participants compared their original portfolio to the AI recommendation in a single forced-choice question with four options: higher expected return and higher risk, higher expected return and lower risk, lower expected return and lower risk, or lower expected return and higher risk. A participant holds a \emph{dominance belief} when choosing the second option, that is, when the participant believes the original portfolio beats the recommendation on both dimensions. Analogous items compare the final portfolio to the recommendation and the final portfolio to the original one. We also compute an objective dominance indicator that equals one when the baseline portfolio has a weakly higher expected return and a weakly lower standard deviation than the recommendation. The survey further records which information participants relied on most, perceived similarity of the task to their real pension decisions, an incentivized financial literacy and cognitive reflection quiz, attitudes toward AI, risk tolerance and patience on zero to ten scales, and standard demographics. Together with comprehension accuracy and screen times recorded during the task, these variables form the controls and secondary measures. The two items that ask about the decision just made, which information participants relied on most and how realistic the task felt, are elicited after the revision and are therefore used only descriptively, never as controls. 


\section{Results}
\label{section:results}

\subsection{Causal Pass-Through of Recommendation Content}
\label{subsec:causal_pass_through}

We begin with a design-based question: how much of the experimentally induced difference between the two recommendation portfolios is transmitted to investors’ final portfolio choices? We refer to this transmission as \textit{pass-through}: the extent to which the difference between the two recommendations appears in the corresponding difference between investors’ subsequent choices. Participants were randomly assigned to receive either an aggressive recommendation, generated by GPT-4 Turbo, or a conservative recommendation, generated by GPT-4o. Relative to the conservative recommendation, the aggressive recommendation offers 3.12 percentage points higher expected return and 6.26 percentage points higher volatility, while having a Sharpe ratio that is 0.072 lower. Because model identity and recommendation content are not independently varied, we interpret the experimental contrast as the causal effect of assignment to these two specific recommendation portfolios, rather than as a general comparison of GPT-4 Turbo and GPT-4o.

\begin{figure}[!ht]
    \centering
    \includegraphics[width=0.80\textwidth]{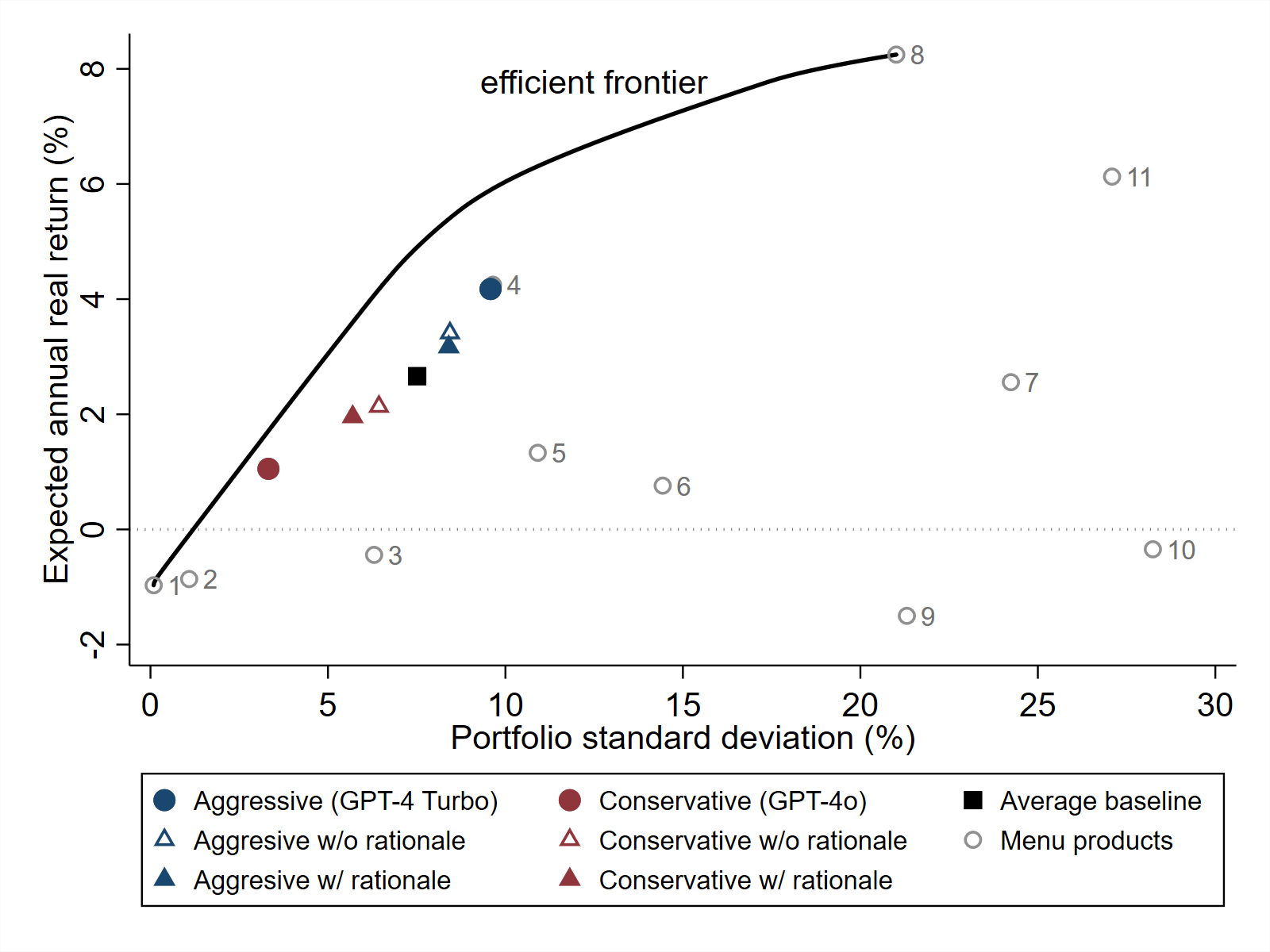}
    \caption{Portfolios and AI Recommendations in Mean--Standard-Deviation Space}
    \label{fig:overall}

    \vspace{0.4em}
    \begin{minipage}{0.90\textwidth}
        \footnotesize
        \textit{Notes:} The figure plots the aggressive and conservative recommendation portfolios, the average baseline portfolio, and average final portfolios in each experimental condition. The aggressive recommendation was generated by GPT-4 Turbo, and the conservative recommendation was generated by GPT-4o. Expected returns are annual real returns, net of 1.2\% expected inflation and a 0.6\% management fee. The solid line is the mean--variance efficient frontier formed from the eleven menu products under long-only weights and uncorrelated returns. Hollow gray circles denote the products, numbered as in the product menu. The black square denotes mean baseline return and risk. Triangles denote average final portfolios: blue for the aggressive recommendation and red for the conservative recommendation, with hollow markers for no rationale and filled markers for a rationale. Filled dots denote the two recommendations. The dotted line marks a zero real return.
    \end{minipage}
\end{figure}

Figure~\ref{fig:overall} provides a visual overview of the experimental contrast in mean--standard-deviation space. The aggressive and conservative recommendations differ sharply in their risk--return profiles, while mean baseline return and risk fall between them. Following the recommendation, average portfolios shift toward the assigned recommendation: investors assigned to the aggressive recommendation move toward higher risk and return, whereas those assigned to the conservative recommendation move in the opposite direction. In both conditions, however, the average final portfolio remains well short of the assigned recommendation. Within each recommendation condition, the final portfolios also appear similar with and without a rationale.

Figure~\ref{fig:overall} provides a transparent economic summary of how the two recommendations shape portfolio risk and return. We next complement this analysis by examining implementation in the full allocation space. Each recommendation specifies weights across eleven products, allowing us to capture product-level reallocations that may not be fully reflected in expected return and standard deviation alone. We therefore construct a measure based on the complete vectors of recommended and chosen portfolio weights.

Let $r^A$ and $r^C$ denote the eleven-dimensional aggressive and conservative recommendation portfolios, respectively. For any portfolio $p$, we define its position on the recommendation axis as

\begin{equation}
a(p) = \frac{ (r^A - r^C) \cdot (p - r^C) }{\| r^A - r^C \|^2}
\label{eq:aggressive_axis}
\end{equation}

\noindent This measure projects portfolio $p$ onto the vector connecting the two recommendations. By construction, $a(r^C)=0$ and $a(r^A)=1$. Consequently, the causal effect of assignment to the aggressive rather than the conservative recommendation on this axis can be interpreted directly as the fraction of the difference between the two recommendation portfolios that passes through to investors' choices.

Our main specification is the following ANCOVA regression:

\begin{equation}
    a(p_{i1})
    =
    \alpha
    + \rho\,a(p_{i0})
    + \beta_A A_i
    + \beta_R R_i \\
    + \beta_{AR}(A_i \times R_i)
    + \varepsilon_i,
\label{eq:axis_ancova}
\end{equation}
where $p_{i0}$ and $p_{i1}$ denote investor $i$'s baseline and final portfolios, respectively. The indicator $A_i$ equals one when the investor is assigned to the aggressive recommendation, and $R_i$ equals one when the recommendation is accompanied by a rationale. To summarize the content effect across the factorial design, we define average pass-through as the aggressive-versus-conservative contrast averaged equally across the two rationale conditions:

\begin{equation}
    \tau_{\mathrm{PT}}
    =
    \beta_A+\frac{1}{2}\beta_{AR}.
    \label{eq:average_pass_through}
\end{equation}

Our headline estimate of average pass-through is 0.368, with a 95\% confidence interval of $[0.303,\,0.433]$ (Table~\ref{tab:axis_pass_through}, Column 2). Assignment to the aggressive recommendation therefore shifts the mean final portfolio by approximately 37\% of the distance separating the two recommendations along the recommendation axis. We strongly reject both no pass-through, $H_0:\tau_{\mathrm{PT}}=0$, and full pass-through, $H_0:\tau_{\mathrm{PT}}=1$, with $p<0.001$ in each case. Recommendation content thus has a substantial causal effect on portfolio choice, but only part of the experimentally induced difference in advice is transmitted to investors' portfolios.

\begin{table}[!ht]
\centering
\footnotesize
\begin{threeparttable}
\caption{Causal Pass-Through of Recommendation Content}
\label{tab:axis_pass_through}
\begin{tabularx}{0.96\textwidth}{@{}>{\raggedright\arraybackslash}p{0.36\textwidth}*{3}{>{\centering\arraybackslash}X}@{}}
\toprule
& \shortstack{(1)\\$a(p_{i1})-a(p_{i0})$}
& \shortstack{(2)\\ANCOVA}
& \shortstack{(3)\\ANCOVA\\+ controls} \\
\midrule
\multicolumn{4}{@{}l}{\emph{Content pass-through}} \\
\addlinespace

Without rationale
    & 0.312*** & 0.324*** & 0.315*** \\
    & (0.050)  & (0.046)  & (0.047)  \\

With rationale
    & 0.433*** & 0.412*** & 0.413*** \\
    & (0.055)  & (0.047)  & (0.047)  \\

Average across rationale conditions
    & 0.372*** & 0.368*** & 0.364*** \\
    & (0.037)  & (0.033)  & (0.033)  \\

\midrule
Controls for baseline axis position
    & No       & Yes      & Yes      \\
Individual controls
    & No       & No       & Yes      \\
Observations
    & 400      & 400      & 400      \\
$R^2$
    & 0.206    & 0.620    & 0.631    \\
\bottomrule
\end{tabularx}
\begin{tablenotes}[flushleft]
\footnotesize
\item \textit{Notes.} The table reports linear combinations of coefficients from regressions with heteroskedasticity-robust standard
errors in parentheses. Column (1) uses the baseline-to-final change in the portfolio's position along the recommendation axis,
$a(p_{i1})-a(p_{i0})$, as the dependent variable. Columns (2) and (3) use the final axis position, $a(p_{i1})$, as the dependent variable and control for the baseline axis position, $a(p_{i0})$. Content pass-through is the aggressive-minus-conservative recommendation contrast within a given rationale condition. Average pass-through assigns equal weight to the conditions with and without a rationale. In each specification, average pass-through is significantly greater than zero and significantly less than one ($p<0.001$ for both tests). Individual controls are age band, gender, comprehension score, an indicator for correctly answering all financial-literacy and cognitive-reflection items, attitudes toward AI, risk tolerance, and patience. $^{***}p<0.01$, $^{**}p<0.05$, $^{*}p<0.10$.
\end{tablenotes}
\end{threeparttable}
\end{table}

Table~\ref{tab:axis_pass_through} shows that the estimated pass-through is highly stable across specifications. Column 1, which uses the baseline-to-final change in the portfolio's position along the recommendation axis, $a(p_{i1})-a(p_{i0})$, yields an average pass-through estimate of 0.372. This estimate means that investors assigned to the aggressive recommendation shift, on average, 0.372 axis units farther in the aggressive direction than those assigned to the conservative recommendation. Because the distance between the two recommendation portfolios is normalized to one, this difference corresponds to 37.2\% of the distance between the recommendations. The baseline-adjusted ANCOVA specification yields the headline estimate of 0.368 (Column 2), and adding the paper's individual-level controls produces an estimate of 0.364 (Column 3). The near-identical estimates show that the main result is not sensitive to whether the outcome is specified as the baseline-to-final change in axis position or as the final axis position conditional on its baseline value, or to the inclusion of demographic, comprehension, attitude-toward-AI, and preference controls.

A rationale could affect portfolio choice in two conceptually distinct ways: it could generate an average shift in investors' positions along the recommendation axis, or it could modify the degree to which recommendation content passes through to their choices. We find no statistically reliable evidence of either effect. In the ANCOVA specification, the rationale effect averaged equally across the aggressive and conservative recommendation conditions, $\beta_R+\frac{1}{2}\beta_{AR}$, is $-0.015$ axis units ($p=0.653$). Estimated pass-through is 0.324 without a rationale and 0.412 with a rationale. Although the point estimate is 0.089 higher with a rationale, this difference is not statistically significant ($p=0.179$). The specifications in Columns 1 and 3 yield the same qualitative conclusion. Thus, the data do not provide statistically reliable evidence that a rationale either systematically shifts portfolio positions along the recommendation axis or modifies the transmission of recommendation content.

Overall, assignment to the aggressive rather than the conservative recommendation produces substantial but incomplete pass-through to investors' portfolio choices. The next subsection examines which portfolio dimensions account for this pass-through, focusing on changes in risk exposure, portfolio composition, concentration, and risk-adjusted performance.


\subsection{Along Which Dimensions Do Investors Follow the Recommendations?}
\label{subsec:dimensions_pass_through}

The preceding subsection establishes that 36.8\% of the experimentally induced difference between the two recommendations passes through to investors' final portfolios. We next examine how this causal pass-through is reflected across economically meaningful dimensions of portfolio choice. Specifically, we estimate the effects of recommendation content on expected return and risk, allocations across risk categories, portfolio breadth and concentration, and risk-adjusted performance.

Let $Y_k(p)$ denote outcome $k$ evaluated for portfolio $p$. The outcomes include expected return, portfolio risk, allocations across risk categories, portfolio breadth and concentration, and risk-adjusted performance. For investor $i$, $Y_k(p_{i0})$ and $Y_k(p_{i1})$ denote the baseline and final values of outcome $k$, respectively. For each outcome, we first measure how much the two recommended portfolios differ:

\begin{equation}
G_k
=
Y_k(r^A)-Y_k(r^C),
\label{eq:outcome_gap}
\end{equation}

\noindent where $r^A$ and $r^C$ denote the aggressive and conservative recommendation portfolios. Thus, $G_k$ is the aggressive-minus-conservative recommendation gap in outcome $k$.

We then estimate the following outcome-specific ANCOVA:

\begin{equation}
Y_k(p_{i1})
=
\alpha_k
+\rho_k Y_k(p_{i0})
+\beta_{Ak}A_i
+\beta_{Rk}R_i
+\beta_{AR,k}(A_iR_i)
+\varepsilon_{ik},
\label{eq:outcome_ancova}
\end{equation}

\noindent where $A_i$ indicates assignment to the aggressive rather than the conservative recommendation, and $R_i$ indicates that the recommendation is accompanied by a rationale. The causal effect of aggressive rather than conservative recommendation content, averaged equally across the two rationale conditions, is

\begin{equation}
\tau_k
=
\beta_{Ak}
+\frac{1}{2}\beta_{AR,k}.
\label{eq:outcome_effect}
\end{equation}

\noindent Finally, for each outcome for which $G_k\neq 0$, we define outcome-specific pass-through as

\begin{equation}
\pi_k
=
\frac{\tau_k}{G_k}.
\label{eq:outcome_pass_through}
\end{equation}

\noindent Hence, $\pi_k$ measures the causal aggressive-versus-conservative difference in investors' final outcome $k$ relative to the corresponding difference between the two recommended portfolios. For example, $\pi_k=0.40$ means that 40\% of the recommendation gap in outcome $k$ passes through to investors' final portfolios. Table~\ref{tab:dimension_pass_through} reports the recommendation gaps, causal effects, and corresponding pass-through estimates.

\begin{table}[!ht]
\centering
\footnotesize
\begin{threeparttable}
\caption{Dimensions of Recommendation Pass-Through}
\label{tab:dimension_pass_through}
\begin{tabularx}{0.98\textwidth}{@{}>{\raggedright\arraybackslash}p{0.38\textwidth}*{3}{>{\centering\arraybackslash}X}@{}}
\toprule
Outcome
    & \shortstack{Advice gap\\$G_k$}
    & \shortstack{Causal effect\\$\tau_k$}
    & \shortstack{Pass-through\\$\pi_k$} \\
\midrule
\multicolumn{4}{@{}l}{\emph{Risk, return, and efficiency}} \\
\addlinespace
Expected return (p.p.)
    & 3.123 & 1.221*** & 0.391*** \\
    &       & (0.133)  & (0.043)  \\
\addlinespace[0.2em]
Standard deviation (p.p.)
    & 6.256 & 2.146*** & 0.343*** \\
    &       & (0.266)  & (0.042)  \\
\addlinespace[0.2em]
Sharpe ratio
    & $-0.071$ & 0.012 & $-0.173$ \\
    &          & (0.012) & (0.175) \\
\addlinespace
\multicolumn{4}{@{}l}{\emph{Allocation by risk category}} \\
\addlinespace
Principal-protected share (p.p.)
    & $-10.0$ & $-3.521$** & 0.352** \\
    &         & (1.437)    & (0.144) \\
\addlinespace[0.2em]
Low-risk share (p.p.)
    & $-20.0$ & $-7.739$*** & 0.387*** \\
    &         & (1.539)     & (0.077) \\
\addlinespace[0.2em]
High-risk share (p.p.)
    & 40.0 & 13.744*** & 0.344*** \\
    &      & (1.659)   & (0.041) \\
\addlinespace[0.2em]
Very-high-risk share (p.p.)
    & 20.0 & 8.083*** & 0.404*** \\
    &      & (1.154)  & (0.058) \\
\addlinespace
\multicolumn{4}{@{}l}{\emph{Portfolio breadth and concentration}} \\
\addlinespace
Number of products held
    & $-3.000$ & $-0.836$*** & 0.279*** \\
    &          & (0.182)     & (0.061) \\
\addlinespace[0.2em]
Herfindahl--Hirschman index
    & 0.055 & 0.003 & 0.055 \\
    &       & (0.011) & (0.201) \\
\addlinespace[0.2em]
Maximum product share (p.p.)
    & 10.0 & 0.701 & 0.070 \\
    &      & (1.163) & (0.116) \\
\bottomrule
\end{tabularx}
\begin{tablenotes}[flushleft]
\footnotesize
\item \textit{Notes.} Each row reports a separate ANCOVA regression of the final outcome on its baseline value, the aggressive-recommendation indicator, the rationale indicator, and their interaction. The causal effect is the aggressive-minus-conservative contrast averaged equally across the two rationale conditions. Pass-through is the causal effect divided by the fixed aggressive-minus-conservative recommendation gap. Heteroskedasticity-robust standard errors are in parentheses; standard errors for pass-through divide the standard error of the causal effect by the absolute recommendation gap. Allocation shares are reported in percentage points. The principal-protected share is the allocation to Product 1; the low-risk share covers Products 1--2, the high-risk share Products 7--11, and the very-high-risk share Products 10--11. The number of products held counts products receiving a strictly positive allocation. All regressions use 400 observations. $^{***}p<0.01$, $^{**}p<0.05$, $^{*}p<0.10$.
\end{tablenotes}
\end{threeparttable}
\end{table}

Column 1 shows that the aggressive recommendation offers 3.123 percentage points more expected return and 6.256 percentage points more portfolio risk than the conservative recommendation, while having a Sharpe ratio that is 0.071 lower. The aggressive recommendation also allocates 10 percentage points less to the principal-protected product and 20 percentage points less to the low-risk category, while allocating 40 percentage points more to the high-risk category and 20 percentage points more to the very-high-risk category. Finally, it contains three fewer products with positive weights and is more concentrated, with a Herfindahl--Hirschman index that is 0.055 higher and a maximum product share that is 10 percentage points higher.

Column 2 shows which of these differences in recommendation content produce detectable causal differences in investors' final portfolios. Assignment to the aggressive recommendation raises expected return by 1.221 percentage points and portfolio standard deviation by 2.146 percentage points. It also reduces the principal-protected and low-risk shares by 3.521 and 7.739 percentage points, respectively, while increasing the high-risk and very-high-risk shares by 13.744 and 8.083 percentage points. All of these effects are in the direction implied by the recommendation gap and are statistically significant. Assignment to the aggressive recommendation also reduces the number of products held by 0.836. By contrast, it does not detectably change investors' final Sharpe ratios, Herfindahl--Hirschman indices, or maximum product shares.

Column 3 places these causal effects on a common scale. Pass-through is 39.1\% for expected return and 34.3\% for portfolio standard deviation. The similarity of these estimates indicates that recommendation content moves portfolios along the risk--return trade-off rather than predominantly through only one of its dimensions. Pass-through across the four allocation categories ranges from 34.4\% to 40.4\%, showing that approximately one-third to two-fifths of the recommended reallocations across risk categories appear in investors' final portfolios. Pass-through in portfolio breadth is somewhat smaller, at 27.9\%. In contrast, the pass-through estimates for the Sharpe ratio, the Herfindahl--Hirschman index, and the maximum product share are statistically indistinguishable from zero. The Sharpe-ratio estimate is negative because the causal effect runs opposite to the recommendation gap, although neither the effect nor the resulting pass-through estimate is statistically significant.

Taken together, the three columns show a clear progression from recommendation content to investor response. The recommendations generate substantial causal changes in portfolio risk, expected return, allocations across risk categories, and breadth, with pass-through generally between approximately 28\% and 40\%. However, the concentration differences embedded in the recommendations do not pass through detectably, and the resulting changes in risk exposure do not produce a detectable difference in risk-adjusted performance.

\begin{figure}[!ht]
    \centering
    \includegraphics[width=0.80\textwidth]{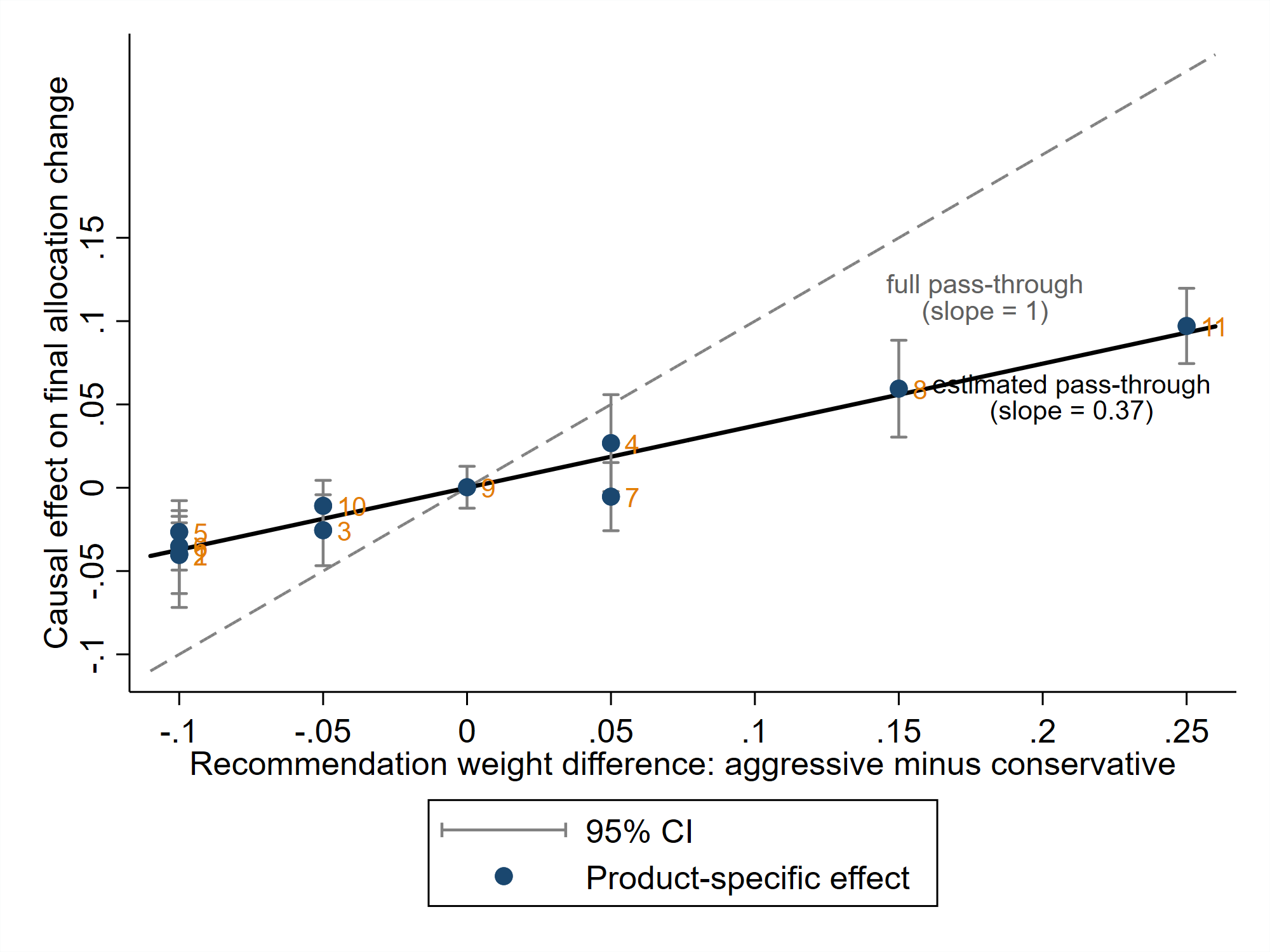}
    \caption{Product-Level Pass-Through of Recommendation Content}
    \label{fig:product_pass_through}

    \vspace{0.4em}
    \begin{minipage}{0.90\textwidth}
        \footnotesize
        \textit{Notes:} Each point represents one of the eleven investment products. The horizontal axis is the product's weight in the aggressive recommendation minus its weight in the conservative recommendation. The vertical axis is the causal effect of assignment to the aggressive rather than the conservative recommendation on the baseline-to-final change in that product's allocation, averaged equally across the two rationale conditions. Portfolio weights are expressed as shares. Vertical bars show heteroskedasticity-robust 95\% confidence intervals. The dashed 45-degree line represents full pass-through. The solid line has slope 0.372, the pooled product-level pass-through estimate; standard errors for the pooled regression are clustered by participant. Product numbers correspond to the product menu.
    \end{minipage}
\end{figure}

Figure~\ref{fig:product_pass_through} decomposes the aggregate result product by product. For product $j$, the horizontal axis reports the recommendation-weight difference, $q_j=r^A_j-r^C_j$, and the vertical axis reports the causal effect of aggressive rather than conservative assignment on the baseline-to-final change in that product's allocation, averaged across rationale conditions. Under full pass-through, each product would lie on the 45-degree line. Instead, the effects are systematically attenuated: nine of the ten products for which the recommendations prescribe different weights have point estimates in the recommended direction, but nearly all lie well inside the full-pass-through benchmark. The largest differences occur for Products 8 and 11, which receive substantially greater weight in the aggressive recommendation, and for Products 1, 2, 5, and 6, which receive greater weight in the conservative recommendation.

The pooled product-level slope is 0.372, exactly matching the change-score estimate in Section~\ref{subsec:causal_pass_through} and closely matching the 0.368 headline ANCOVA estimate. The exact correspondence with the change-score estimate is not accidental: the recommendation-axis measure is the projection of the eleven product-level allocation changes onto the vector connecting the two recommendations. Appendix Figure~\ref{fig:groups_1_4} reports the raw baseline and final allocations separately for all four treatment cells.

The answer to the subsection's question is therefore clear. Investors follow the recommendations primarily by changing their portfolios' expected return, volatility, exposure to different risk categories, product composition, and breadth. They do not simply reproduce every feature of the recommended portfolios: recommendation content does not detectably pass through to the Sharpe ratio or to standard measures of portfolio concentration. The next subsection examines how individual investors implement these changes, separating the direction and magnitude of their revisions.


\subsection{How Do Investors Follow the Recommendations?}
\label{subsec:individual_implementation}

The preceding subsections establish substantial causal pass-through of recommendation content and identify the portfolio dimensions through which it occurs. We next turn from what changes in investors' portfolios to how investors implement the assigned recommendations. Aggregate pass-through can reflect several patterns of individual adjustment: widespread partial movement toward the recommendation, full adoption by some investors and little adjustment by others, or movement toward the recommendation accompanied by changes along other portfolio dimensions. To characterize this implementation process, we examine three distinct margins: whether investors revise their portfolios at all, whether their revisions are directed toward the assigned recommendation, and how much of the recommended move they implement.

As a descriptive starting point, the mean Euclidean distance between an investor's portfolio and the received recommendation falls from 0.52 at baseline to 0.41 after advice. The mean decline of 0.11 is statistically significant ($p<0.001$) and represents a 21\% reduction. This within-investor decline is not the causal pass-through estimate from Section~\ref{subsec:causal_pass_through}: every investor receives a recommendation, and Euclidean distance is a different outcome from position on the recommendation axis. Nevertheless, the decline is advice-specific. Distance to the recommendation not received falls by only 0.032, so distance to the assigned recommendation declines by an additional 0.078 ($p<0.001$). Some decline toward the unassigned recommendation is expected because the directions from investors' baseline portfolios toward the two recommendations are positively aligned, with a mean cosine of 0.70.

Distance provides an overall summary of how closely investors' final portfolios align with their assigned recommendations. We complement this measure by examining the investor-level revisions that produce this alignment. This approach distinguishes nonrevision from active adjustment and, among active revisions, separates movement in the direction of the recommendation from simultaneous movement along other portfolio dimensions. Let $p_{i0}$ and $p_{i1}$ denote investor $i$'s baseline and final portfolios, respectively, and let $r_i$ denote the assigned recommendation. Define the portfolio revision as $\Delta_i=p_{i1}-p_{i0}$ and the advice gap as $g_i=r_i-p_{i0}$. We decompose the revision into its projection onto the advice gap and an orthogonal remainder:

\begin{equation}
    \underbrace{p_{i1}}_{\text{final}}
    =
    \underbrace{p_{i0}}_{\text{baseline}}
    +
    \underbrace{\beta_i(r_i-p_{i0})}_{\text{AI-directed}}
    +
    \underbrace{\eta_i}_{\text{other}},
    \qquad
    \eta_i \perp (r_i-p_{i0}).
    \label{eq:revision_decomposition}
\end{equation}

The projection coefficient $\beta_i = \langle \Delta_i, g_i\rangle / \lVert g_i\rVert^{2}$ is the weight on advice, the multi-asset analogue of the corresponding scalar measure in the advice-taking literature \citep[e.g.,][]{harvey1997taking,yaniv2000advice}. It measures the fraction of the recommended move the participant implements. The directedness $\cos\theta_i = \langle \Delta_i, g_i\rangle / (\lVert \Delta_i\rVert\,\lVert g_i\rVert)$ measures how much of the revision's direction is aimed at the recommendation, independent of how far the participant moves. It equals 1 for a revision aimed exactly at the recommendation, 0 for an orthogonal revision, and is negative for movement away. The identity $\beta_i = \cos\theta_i \times (\lVert \Delta_i\rVert / \lVert g_i\rVert)$ links the two, so the weight on advice is directedness times relative step size.

\begin{figure}[!ht]
\centering
\includegraphics[width=0.80\textwidth]{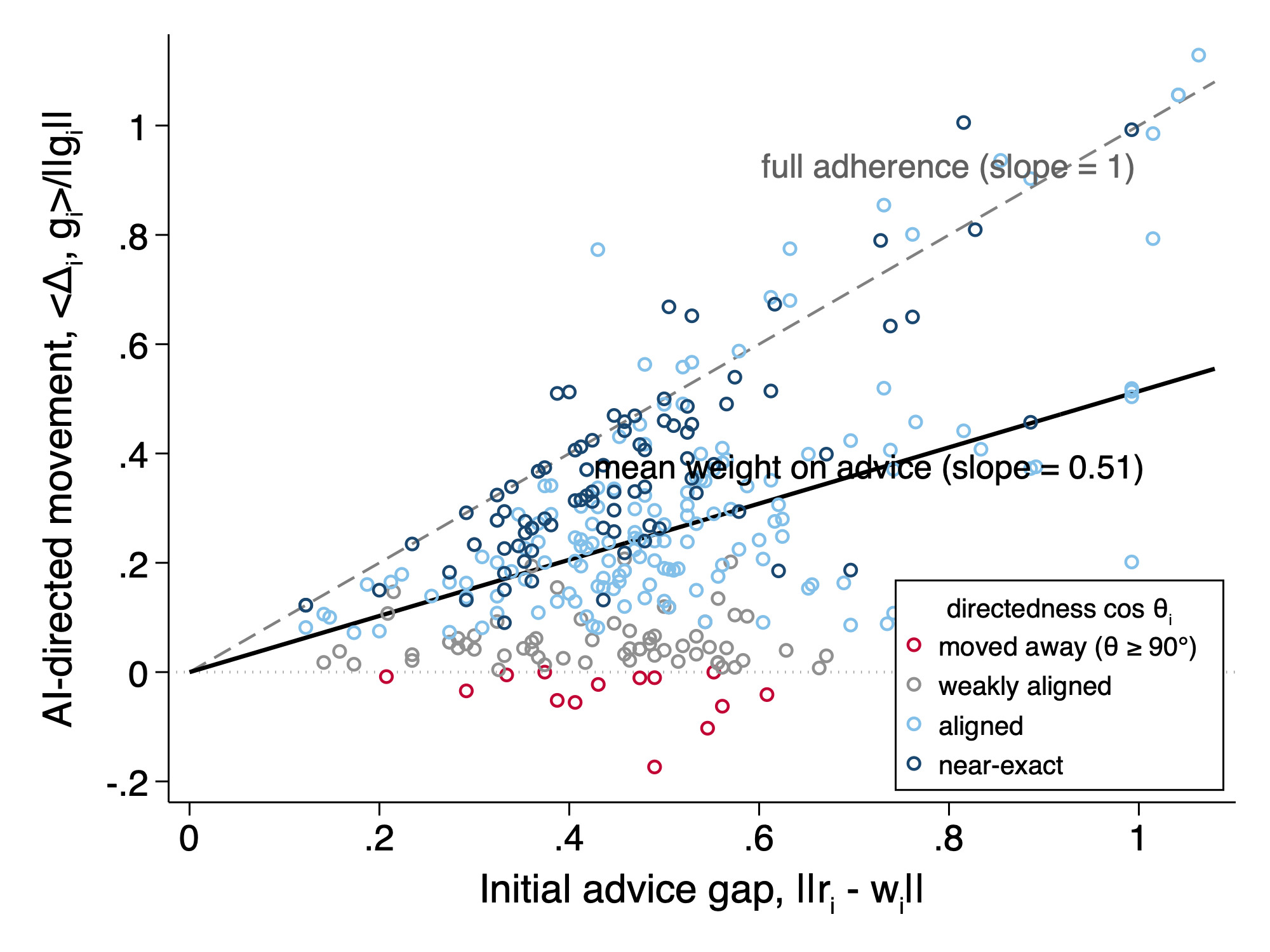}
\caption{Direction, Magnitude, and Weight on Advice}
\label{fig:directedness}

\vspace{0.4em}
\begin{minipage}{0.90\textwidth}
    \footnotesize
    \textit{Notes:} Each point represents one of the 323 investors who revised their allocation. The 77 non-revisers are excluded because directedness is undefined when the revision is zero; their weight on advice is zero by construction. The horizontal axis reports the initial advice gap, $\lVert r_i-p_{i0}\rVert$. The vertical axis reports AI-directed movement, $\langle\Delta_i,g_i\rangle/\lVert g_i\rVert$. The dashed 45-degree line represents full implementation ($\beta_i=1$), and the solid line has slope 0.51, the mean weight on advice among revisers. Points are shaded by directedness, $\cos\theta_i$.
\end{minipage}

\end{figure}

Of the 400 participants, 323 (81 percent) revise their allocation. Among revisers, the revision is strongly aimed at the recommendation. The mean directedness is $\cos\theta = 0.65$ (median 0.76), and 95 percent of revisers move toward the received recommendation, while 4 percent move away and three revise in a direction orthogonal to the advice. In magnitude, however, participants systematically under-adjust. The mean weight on advice is $\bar\beta = 0.51$ (median 0.52), so a typical participant implements about half of the recommended move. Revisers thus aim in roughly the right direction but take a step only about three quarters as long as the advice gap. The orthogonal remainder $\eta_i$ accounts on average for close to half of the squared length of the revision. This is why the fall in distance understates adherence. \autoref{fig:directedness} shows this pattern. The full distribution of $\cos\theta_i$ appears in \autoref{fig:costheta_hist} in the Appendix. 

Figure~\ref{fig:directedness} displays this decomposition investor by investor. The horizontal axis is the initial advice gap, $\lVert g_i\rVert$, and the vertical axis is the signed length of the revision projected onto the recommendation direction, $\langle\Delta_i,g_i\rangle/\lVert g_i\rVert=\beta_i\lVert g_i\rVert$. A point on the dashed 45-degree line represents full implementation. Most observations lie above zero but below this benchmark: investors generally move in the recommended direction but only partway. The solid line, with slope 0.51, summarizes the mean weight on advice among revisers. The dispersion around it shows substantial heterogeneity in both direction and magnitude.

The individual-level weight on advice and the aggregate pass-through estimate answer related but distinct questions. Assigning zero weight to the 77 non-revisers gives a full-sample mean weight on advice of 0.41, compared with the 0.51 mean among revisers. Neither mean is expected to equal the 0.368 aggregate pass-through estimate in Section~\ref{subsec:causal_pass_through}. The weight on advice scales each investor's revision by that investor's own distance and direction to the assigned recommendation, whereas aggregate pass-through is a between-treatment causal contrast projected onto the fixed axis connecting the two recommendations. Together, the estimates show that incomplete aggregate transmission reflects both an extensive margin---nearly one fifth of investors do not revise---and an intensive margin---active revisers implement only part of the recommended move and devote some of their rebalancing to directions orthogonal to the advice.

These implementation patterns do not differ detectably across the experimental conditions. Appendix Table~\ref{tab:distance_reg} shows that neither recommendation type, the rationale, nor their interaction significantly predicts final distance conditional on baseline distance or the change in distance. The change-score specification explains little of the cross-sectional variation in adjustment ($R^2=0.045$). Thus, recommendation content determines the direction in which portfolios move, but it does not explain much of the heterogeneity in how completely individual investors implement the advice. The next subsection asks whether this heterogeneity reflects the congruence between the randomly assigned recommendation and the investor's baseline portfolio.


\subsection{Drivers of Adherence: Overconfidence and Baseline Inefficiency}
\label{subsec:drivers_adherence}

The preceding subsection shows that incomplete adherence reflects distinct margins of individual behavior: whether investors revise, how accurately their revisions point toward the recommendation, and how far they move. We now examine two investor characteristics that may account for this heterogeneity. The first is a subjective assessment of the recommendation relative to one's own portfolio, and the second is the objective efficiency of the investor's baseline portfolio. These characteristics predict different margins of implementation.

The identification status of these analyses differs from that of the preceding treatment comparisons. Recommendation content and the provision of a rationale are randomly assigned, whereas investors' assessments of their own portfolios and the efficiency of their baseline portfolios are not. Moreover, the subjective comparison is elicited after the portfolio decision. The estimates in this subsection should therefore be interpreted as descriptive associations rather than causal effects. In particular, we cannot rule out the possibility that an investor who chose not to follow the advice subsequently reported a more favorable assessment of the baseline portfolio.

We first consider whether investors who are excessively confident in their own portfolios follow the recommendation less closely. In the post-task survey, participants compare the expected return and risk of their baseline portfolio with those of the recommendation they received. We define the indicator $OC_i$ to equal one if investor $i$ believes that the baseline portfolio dominates the recommendation---that it has both a higher expected return and lower risk---although the computed portfolio moments show that it does not. This indicator captures \emph{overconfidence} in the sense of overestimation.\footnote{In the taxonomy of \citet{moore2008trouble}, overestimation is an overly favorable assessment of one's own performance, overplacement is the belief that one ranks above others, and overprecision is excessive certainty about the accuracy of one's information. Our measure marks investors who overestimate the performance of the portfolio they chose, believing it beats the received recommendation on both dimensions when it does not. Because the benchmark is the specific recommendation received, the belief also has an element of overplacement relative to the advisor. It is distinct from the overprecision notion that underlies classic models of overconfident trading, in which investors overstate the precision of their private information \citep{odean1998volume}. Overprecision also predicts behavior outside financial markets \citep{ortoleva2015overconfidence}.} This dominance belief is common and almost always incorrect. 27\% of participants believe that their baseline portfolio dominates the recommendation on both dimensions, whereas only 3\% actually hold such a portfolio. The discrepancy appears under both recommendations. Six percent of investors assigned to the GPT-4 Turbo recommendation actually dominate it, and none of those assigned to the GPT-4o recommendation do, yet more than one quarter of each group believes that they do.

Let $d_i^0=\lVert r_i-p_{i0}\rVert$ and $d_i^1=\lVert r_i-p_{i1}\rVert$ denote investor $i$'s baseline and final distances from the assigned recommendation. We estimate

\begin{equation}
d_i^0-d_i^1
=
\alpha
+\beta OC_i
+\gamma d_i^0
+\delta_1 GPT4o_i
+\delta_2 Rationale_i
+\delta_3(GPT4o_i\times Rationale_i)
+X_i'\theta
+\varepsilon_i,
\label{eq:overconfidence_gap}
\end{equation}

\noindent where $d_i^0-d_i^1$ is the gap closed toward the received recommendation, $GPT4o_i=1-A_i$, and $X_i$ contains the individual controls used throughout the analysis. A larger dependent variable indicates greater adherence. Conditional on baseline distance, equation~\eqref{eq:overconfidence_gap} is a reparameterization of the final-distance specification reported in Appendix Table~\ref{tab:distance_reg}.

\begin{table}[!ht]
\centering
\footnotesize
\begin{threeparttable}
\caption{Drivers of Adherence: Overconfidence and Baseline Inefficiency}
\label{tab:drivers_adherence}
\begin{tabularx}{0.92\textwidth}{@{}>{\raggedright\arraybackslash}p{0.53\textwidth}*{3}{>{\centering\arraybackslash}X}@{}}
\toprule
& (1) & (2) & (3) \\
\midrule
\multicolumn{4}{@{}l}{\emph{Panel A. Gap closed toward the recommendation, $d_i^0-d_i^1$}} \\
\addlinespace
Overconfident (dominance)
& $-0.049^{***}$ & $-0.036^{**}$ & $-0.056^{**}$ \\
& (0.015)        & (0.016)        & (0.024)        \\
\addlinespace
Overconfidence $\times$ treatment interactions & No & No & Yes \\
Individual controls & No & Yes & Yes \\
Baseline distance & Yes & Yes & Yes \\
Treatment indicators & Yes & Yes & Yes \\
Observations & 400 & 400 & 400 \\
$R^2$ & 0.040 & 0.065 & 0.068 \\
\addlinespace[0.6em]
\multicolumn{4}{@{}l}{\emph{Panel B. Revision magnitude, $\lVert p_{i1}-p_{i0}\rVert$}} \\
\addlinespace
Bad ($\mathrm{Sharpe}_{i0}<\mathrm{Sharpe}_{r_i}$) & $0.045^{**}$ & 0.065 &  \\
& (0.022)      & (0.041) &  \\
\addlinespace
Inefficiency $\times$ treatment interactions & No & Yes &  \\
Individual controls & Yes & Yes &  \\
Baseline distance & Yes & Yes &  \\
Treatment indicators & Yes & Yes &  \\
Observations & 400 & 400 &  \\
$R^2$ & 0.072 & 0.074 &  \\
\bottomrule
\end{tabularx}
\begin{tablenotes}[flushleft]
\footnotesize
\item \textit{Notes.} Robust standard errors are in parentheses. Panel A estimates equation~\eqref{eq:overconfidence_gap}. Overconfident (dominance) equals one if the investor believes that the baseline portfolio has both a higher expected return and lower risk than the received recommendation although the computed portfolio moments show that it does not. Panel B estimates equation~\eqref{eq:inefficiency_revision}. Bad equals one if the Sharpe ratio of the baseline portfolio is below that of the received recommendation. Treatment indicators include the GPT-4o recommendation indicator, the rationale indicator, and their interaction. The treatment-interaction specifications additionally interact the investor characteristic with GPT-4o and the rationale. Both panels use the full sample, including investors who do not revise. Full coefficient estimates appear in Appendix Table~\ref{tab:drivers_full}. $^{***}p<0.01$, $^{**}p<0.05$, $^{*}p<0.10$.
\end{tablenotes}
\end{threeparttable}
\end{table}

Panel A of Table~\ref{tab:drivers_adherence} shows that dominance-overconfident investors close significantly less of the gap. Without individual controls, the coefficient on $OC_i$ is $-0.049$ ($p<0.01$). It remains negative and significant after adding the full control set, including financial literacy and comprehension: the estimate is $-0.036$ ($p<0.05$). The relationship therefore does not appear to reflect lower measured competence among overconfident investors. Descriptively, these investors close approximately 14\% of their initial distance to the recommendation, compared with 26\% among other investors. Column (3) allows the association to differ across the experimental conditions. The estimated effect in the reference treatment is $-0.056$ ($p<0.05$), while the interactions with recommendation type and the rationale are jointly insignificant ($p=0.45$). The negative relationship between overconfidence and adherence is therefore not concentrated in a particular model or rationale condition.

This difference is a following gap, not an engagement gap. When revision magnitude replaces the gap closed as the dependent variable, the coefficient on $OC_i$ is 0.011 ($p=0.73$). Overconfident investors revise their portfolios by approximately the same amount as other investors, but direct less of that movement toward the recommendation. Thus, overconfidence is associated with the direction of the revision rather than the willingness to reconsider the portfolio.

The amount responds to a different margin, the quality of the starting point. We define an indicator $\mathrm{Bad}_i$ that equals one if the Sharpe ratio of participant $i$'s baseline portfolio is below the Sharpe ratio of the received recommendation, which holds for 59\% of participants, and estimate

\begin{equation}
\mathrm{Revision}_i
=
\alpha
+\beta Bad_i
+\gamma d_i^0
+\delta_1 GPT4o_i
+\delta_2 Rationale_i
+\delta_3(GPT4o_i\times Rationale_i)
+X_i'\theta
+\varepsilon_i,
\label{eq:inefficiency_revision}
\end{equation}
where $\mathrm{Revision}_i \equiv \lVert p_{i1}-p_{i0}\rVert$. Because the two recommendations have different Sharpe ratios, $Bad_i$ depends partly on which recommendation an investor receives. All specifications therefore condition on the treatment indicators.

Panel B shows that investors whose baseline portfolios are less efficient than the received recommendation revise more. The coefficient on $Bad_i$ is 0.045 ($p<0.05$), approximately 16\% of the average revision magnitude of 0.28. Revision magnitude also increases with baseline distance, as shown by the full estimates in Appendix Table~\ref{tab:drivers_full}.

Column (2) interacts $\mathrm{Bad_i}$ with recommendation type and the rationale. The estimated reference-treatment effect rises to 0.065 but is no longer statistically significant once the association is divided across the treatment cells. The interaction coefficients are small and statistically indistinguishable from zero. The evidence therefore does not suggest that the association between initial inefficiency and revision magnitude is specific to a particular recommendation or to the provision of a rationale.

Larger revisions by initially inefficient investors do not necessarily produce closer adherence. When the gap closed replaces revision magnitude as the dependent variable in equation~\eqref{eq:inefficiency_revision}, the coefficient on $Bad_i$ is 0.001, with a standard error of 0.015. Initially inefficient investors move more, but they do not end significantly closer to the recommendation.

The individual decomposition introduced in Section~\ref{subsec:individual_implementation} sharpens this distinction. Among the 323 investors who revise, dominance-overconfident investors have mean directedness of 0.54, compared with 0.69 among other revisers. Conditional on baseline distance, treatments, and individual controls, overconfidence is associated with a 0.108 reduction in directedness (standard error 0.042), while its association with revision magnitude is only 0.010 (standard error 0.027). By contrast, revisers whose baseline portfolios are Sharpe-dominated place 0.131 more weight on the advice (standard error 0.040), while their directedness differs by only 0.044 (standard error 0.038).

These results identify a division between two margins of implementation. Subjective overconfidence is associated with poorer aim: investors revise, but less of their revision points toward the recommendation. Objective baseline inefficiency is associated with a larger step: investors revise more, without aiming more accurately or necessarily ending closer to the recommendation. Because these investor characteristics are not randomly assigned, this evidence should not be interpreted causally. It nevertheless shows that heterogeneity in adherence cannot be summarized by a single measure of responsiveness: the direction and magnitude of portfolio revisions are distinct behavioral margins associated with different investor characteristics.


\subsection{Robustness Checks}
\label{subsec:heterogeneity_robustness}

We conclude the results by assessing robustness across participant subgroups, outcome definitions, and estimation approaches. Across these checks, recommendation content continues to shift portfolio risk and expected return in the direction of the assigned portfolio, while adherence and risk-adjusted performance remain similar across model and rationale conditions.

Appendix Table~\ref{tab:hetero_bin} examines heterogeneity by gender and by two post-task self-reports: whether participants ranked the GPT recommendation as their most important information source and whether the task felt similar to an actual pension decision. All twelve subgroup-by-treatment interactions are statistically indistinguishable from zero. Some of the post-task subgroup indicators are associated with portfolio behavior, but these correlations are descriptive because the indicators were elicited after the revision. Importantly, none alters the randomized comparisons across recommendation-type or rationale conditions.

A placebo test provides direct evidence that pass-through tracks the recommendation an investor actually receives. At the investor-product level, allocation changes load strongly on the gap between the baseline portfolio and the assigned recommendation: $b_{\mathrm{own}}=0.369$ with a standard error of $0.025$. By contrast, the loading on the recommendation generated by the model not assigned to the investor is $b_{\mathrm{other}}=0.001$ with a standard error of $0.017$ ($p=0.94$); the two coefficients differ at $p<0.001$. This result closely matches the aggregate pass-through estimates and confirms that participants respond to the specific recommendation they receive rather than merely rebalancing in a direction common to both recommendations.

The main conclusions are stable across outcome definitions and estimators. The full-factorial change-score specifications in Appendix Table~\ref{tab:treat_summary} show that recommendation type predicts changes in expected return and volatility, the dimensions along which the recommendations differ most sharply. Recommendation type, the rationale, and their interaction do not detectably predict changes in distance, the Sharpe ratio, or revision magnitude. The few marginal rationale terms for expected return and volatility do not translate into a detectable Sharpe-ratio improvement. Individual fixed-effects estimates among the 323 active revisers in Appendix Table~\ref{tab:fe_time_rationale} yield the same pattern: GPT-4 Turbo shifts portfolios toward higher risk and return, GPT-4o shifts them toward lower risk and return, and neither significantly improves the Sharpe ratio at the 5\% level.

Finally, investors with below-median baseline Sharpe ratios experience larger subsequent improvements than those starting above the median (Appendix Table~\ref{tab:sharpe_eff}). Because the experiment contains no group that revises without receiving advice, this pattern is best interpreted as convergence from different starting points rather than as a causal efficiency gain from AI advice. Power calculations further show that the treatment specifications can detect effects of approximately $0.06$--$0.08$ in distance and the Sharpe ratio and $0.10$--$0.15$ in revision magnitude at 80\% power. The null results therefore rule out advice-side effects approaching the magnitude of the average behavioral response, although smaller effects remain possible.

Taken together, these checks reinforce the central result: investors respond specifically and substantially to recommendation content, but the completeness and efficiency of implementation vary little across model, rationale, and observable subgroup conditions.


\section{Conclusion}
\label{section:conclusion}

We study how AI-generated recommendation content is transmitted into individual economic decisions. In an experiment with 400 employed South Korean adults enrolled in workplace defined contribution pension plans, participants revise an initial portfolio after receiving either an aggressive or a conservative AI-generated recommendation. Approximately 37 percent of the experimentally induced difference between these recommendations passes through to final portfolios. This transmission appears in expected return, volatility, exposure to different risk categories, and portfolio breadth. It does not produce a detectable difference in risk-adjusted performance. AI advice therefore changes the level and composition of risk participants assume, even when it does not improve how efficiently they bear that risk.

Individual behavior helps explain why pass-through is incomplete. Eighty-one percent of participants revise their portfolios, and 95 percent of revisers move in a direction aligned with the recommendation. Yet the typical reviser implements only about half of the recommended adjustment and also makes changes orthogonal to the advice. Participants therefore neither accept nor reject AI advice wholesale; they combine it with their own prior judgments. The descriptive heterogeneity results further distinguish two margins of this process. Participants who mistakenly believe that their initial portfolio dominates the recommendation revise by as much as others but direct less of their revision toward the advice. Participants whose initial portfolios are less efficient revise more, but do not end significantly closer to the recommendation.

These findings have implications for organizations deploying AI-based financial guidance and policymakers regulating such services. Two frontier AI models given identical information produced substantively different recommendations, and participants followed the recommendations they received, albeit incompletely. As a result, organizations that deploy such systems may intentionally or unintentionally steer users toward particular risk-return profiles, while the particular AI systems to which retail investors turn may introduce unexpected yet consequential variation in their portfolios and, potentially, in financial markets. However, our evidence is limited to a one-time, nonbinding, hypothetical portfolio-choice environment, and the observed relationships between participant characteristics and adherence are descriptive rather than causal. Therefore, further research is called for to identify which features of users and recommendations determine whether, how, and to what extent people follow AI-generated advice.

\section*{Data and Code Availability}

A repository for the replication materials has been created on the 
\href{https://osf.io/gufb7/overview?view_only=8bcbcfb0717f47248f87518f664628cf}{Open Science Framework (OSF)}.
The replication package will be made publicly available upon publication.



\newpage

\bibliographystyle{apalike}
\bibliography{references} 

\clearpage

\newpage

\appendix

\numberwithin{table}{section}
\numberwithin{figure}{section}
\renewcommand{\thetable}{\thesection.\arabic{table}}
\renewcommand{\thefigure}{\thesection.\arabic{figure}}


\section{Additional Tables}
\label{appendix:additional_tables}

\subsection{Balance Test}\label{appendix:balance}

Table \ref{tab:balance} reports, for each of the four treatment cells, the mean of the baseline portfolio moments and participant characteristics, together with the $p$-value of a joint $F$-test that the four cell means are equal. Baseline portfolio moments and demographics are statistically indistinguishable across cells. Age and gender are balanced by design through quota sampling. The baseline distance to the recommendation is not listed because it is defined relative to the received recommendation and therefore differs across model treatments by construction. Within each model treatment it is balanced across the rationale conditions ($p=0.73$ for GPT-4 Turbo and $p=0.89$ for GPT-4o).

\begin{table}[!ht]
\centering
\footnotesize
\begin{threeparttable}
\caption{Balance across treatment cells}\label{tab:balance}
\begin{tabularx}{\textwidth}{@{}l*{5}{>{\centering\arraybackslash}X}@{}}
\toprule
 & GPT-4 Turbo & GPT-4 Turbo+R & GPT-4o & GPT-4o+R & $p$-value \\
\midrule
Baseline expected return      & 4.671 & 4.285 & 4.334 & 4.555 & 0.593 \\
Baseline standard deviation   & 7.965 & 7.346 & 7.627 & 7.121 & 0.629 \\
Baseline Sharpe ratio         & 0.473 & 0.454 & 0.484 & 0.511 & 0.217 \\
Comprehension score           & 2.340 & 2.280 & 2.320 & 2.460 & 0.322 \\
Financial literacy and CRT    & 6.490 & 6.560 & 6.680 & 6.910 & 0.598 \\
AI attitude                   & 50.150 & 50.580 & 51.490 & 50.690 & 0.859 \\
Risk tolerance                & 5.050 & 5.190 & 5.120 & 5.070 & 0.974 \\
Patience                      & 5.740 & 6.180 & 6.400 & 6.080 & 0.120 \\
Education                     & 3.850 & 4.020 & 3.920 & 3.880 & 0.378 \\
Female                        & 0.500 & 0.500 & 0.500 & 0.500 & 1.000 \\
Age band                      & 1.500 & 1.500 & 1.500 & 1.500 & 1.000 \\
\midrule
Observations (per cell)       & 100 & 100 & 100 & 100 & \\
\bottomrule
\end{tabularx}
\begin{tablenotes}[flushleft]
\footnotesize
\item \textit{Notes.} Each entry is a treatment-cell mean, with 100 participants per cell and 400 in total. The last column reports the $p$-value of a joint $F$-test that the four cell means are equal, with heteroskedasticity-robust standard errors. Expected return and standard deviation are annualized and in percent. The financial literacy and CRT score is the number of the eleven quiz items answered correctly, reported here as a score although the regressions control for an indicator that all eleven are correct. Risk tolerance and patience are on zero to ten scales. Age band is coded 1 for ages 35 to 45 and 2 for ages 46 to 55. Age and gender are fixed by the sampling quota, so their cell means are equal by construction.
\end{tablenotes}
\end{threeparttable}
\end{table}

\subsection{Within-Person Fixed-Effects Estimates}
Table \ref{tab:fe_time_rationale} exploits the panel structure by estimating within-person changes in portfolio \emph{levels} using individual fixed effects (FE) among the 323 participants who actively revised their portfolios. Stacking each participant's baseline ($t=0$) and post-recommendation ($t=1$) observations, we estimate, separately by GPT version and outcome $Y\in\{E,\mathrm{Std},\mathrm{Sharpe}\}$,
\begin{equation}
Y_{it} = \mu_i + \phi\,\mathrm{Post}_{it} + \psi\,(\mathrm{Post}\times\mathrm{Rationale})_{it} + \varepsilon_{it},
\label{eq:panelfe}
\end{equation}
where $\mu_i$ are individual fixed effects and $\mathrm{Post}_{it}=1$ for the post-recommendation observation. The key regressor is \textit{Post}, so its coefficient $\phi$ can be interpreted as the average within-person shift from baseline to the post-recommendation portfolio for the no-rationale group within each GPT version. The interaction term \textit{Post}$\times$Rationale ($\psi$) captures how this within-person shift differs when a rationale is provided.

Within the GPT-4 Turbo subsample (Columns (1) to (3)), revisers increase both expected return and risk after receiving advice: the \textit{Post} coefficient is positive for $E$ and $\mathrm{Std}$. The Sharpe estimate is positive and marginally significant. The Post$\times$Rationale interaction is statistically indistinguishable from zero across outcomes, providing little evidence that the rationale systematically changes within-person post shifts under GPT-4 Turbo.

Within the GPT-4o subsample (Columns (4) to (6)), revisers move in the opposite direction: the post indicator is negative for both $E$ and $\mathrm{Std}$ and is precisely estimated, indicating reductions in both expected return and risk after advice. Sharpe again shows no systematic post shift: the estimated post coefficient is close to zero and statistically indistinguishable from zero. The most notable rationale effect in this panel specification appears for expected return under GPT-4o: the Post$\times$Rationale coefficient is negative and statistically significant, suggesting that among GPT-4o revisers, adding a rationale is associated with an additional within-person reduction in expected return. However, this does not translate into an improvement in Sharpe, and the corresponding risk response to rationale is not precisely estimated.

\begin{table}[!ht]
\centering
\begin{threeparttable}
\caption{Panel FE estimates: within-person changes by rationale}
\label{tab:fe_time_rationale}
\footnotesize
\begin{tabularx}{\textwidth}{@{}l*{6}{>{\centering\arraybackslash}X}@{}}
\toprule
 & \multicolumn{3}{c}{GPT-4 Turbo} & \multicolumn{3}{c}{GPT-4o} \\
\cmidrule(lr){2-4}\cmidrule(lr){5-7}
 & (1) $E$ & (2) Std & (3) Sharpe & (4) $E$ & (5) Std & (6) Sharpe \\
\midrule
Post                                 & 0.700*** & 0.615*   & 0.037*   & -0.493*** & -1.470*** & 0.022 \\
                                     & (0.198)  & (0.336)  & (0.021)  & (0.145)   & (0.390)   & (0.014) \\
Post $\times$ Rationale               & 0.105    & 0.629    & 0.006    & -0.507**  & -0.309    & -0.023 \\
                                     & (0.278)  & (0.458)  & (0.029)  & (0.231)   & (0.550)   & (0.021) \\
\midrule
Observations                          & 324       & 324       & 324       & 322        & 322        & 322 \\
Individuals (id)                      & 162       & 162       & 162       & 161        & 161        & 161 \\
Within $R^2$                           & 0.156     & 0.106     & 0.047     & 0.226      & 0.181      & 0.014 \\
Individual FE                         & Yes       & Yes       & Yes       & Yes        & Yes        & Yes \\
\bottomrule
\end{tabularx}
\begin{tablenotes}[flushleft]
\footnotesize
\item \textit{Notes.} Fixed-effects (within) regressions with individual fixed effects. Robust standard errors clustered at the individual level in parentheses. The sample is the 323 participants who revised their allocation, split by model treatment into 162 and 161. The 77 non-revisers are excluded because their baseline and post-recommendation portfolios are identical, so they contribute no within-person variation. $^{***}p<0.01$, $^{**}p<0.05$, $^{*}p<0.10$.
\end{tablenotes}
\end{threeparttable}
\end{table}

\subsection{Heterogeneity by Participant Characteristics}

Table \ref{tab:hetero_bin} reports subgroup-by-treatment interactions for gender, whether the participant treated the GPT recommendation as the most important (\textit{GPT first}), and whether the task felt realistic, on the change in distance and on revision magnitude. For a subgroup indicator $G_i$ and outcome $Y\in\{\Delta d_i,\,\mathrm{revision}_i\}$ we estimate
\begin{align}
Y_i &= \alpha + \lambda_0\,G_i + \gamma_1\,\mathrm{GPT-4o}_i + \gamma_2\,\mathrm{Rationale}_i + \gamma_3\,(\mathrm{GPT-4o}\times\mathrm{Rationale})_i \notag \\
&\quad + \lambda_1\,(G_i\times\mathrm{GPT-4o}_i) + \lambda_2\,(G_i\times\mathrm{Rationale}_i) + \mathbf{X}_i'\boldsymbol{\theta} + \varepsilon_i,
\label{eq:hetero}
\end{align}
where the subgroup main effect $\lambda_0$ is estimated directly, since \textit{GPT first} and \textit{realistic task} are elicited after the revision and are therefore not among the controls. We report the interaction terms $\lambda_1$ and $\lambda_2$. They are generally small and imprecise. Because \textit{GPT first} and \textit{realistic task} are post-revision self-reports, splitting on them breaks the randomization within subgroups, so these results are descriptive only.

\begin{table}[!ht]
\centering
\begin{threeparttable}
\caption{Heterogeneity by participant characteristics}
\label{tab:hetero_bin}
\footnotesize
\begin{tabularx}{\textwidth}{@{}l*{6}{>{\centering\arraybackslash}X}@{}}
\toprule
 & \multicolumn{2}{c}{Female} & \multicolumn{2}{c}{GPT first} & \multicolumn{2}{c}{Realistic} \\
\cmidrule(lr){2-3}\cmidrule(lr){4-5}\cmidrule(lr){6-7}
 & (1) $\Delta d_i$ & (2) Rev. & (3) $\Delta d_i$ & (4) Rev. & (5) $\Delta d_i$ & (6) Rev. \\
\midrule
Subgroup                    & -0.007 & -0.063 & -0.229*** & 0.140** & 0.056** & -0.041 \\
                            & (0.024) & (0.047) & (0.035) & (0.054) & (0.026) & (0.043) \\
Subgroup $\times$ GPT-4o    & -0.026 & 0.072 & -0.010 & 0.058 & -0.009 & -0.004 \\
                            & (0.029) & (0.050) & (0.057) & (0.073) & (0.031) & (0.051) \\
Subgroup $\times$ rationale & 0.041 & 0.038 & 0.029 & 0.004 & -0.001 & -0.052 \\
                            & (0.030) & (0.052) & (0.056) & (0.071) & (0.030) & (0.050) \\
\midrule
Treatment indicators        & Yes & Yes & Yes & Yes & Yes & Yes \\
Controls                    & Yes & Yes & Yes & Yes & Yes & Yes \\
Observations                & 400 & 400 & 400 & 400 & 400 & 400 \\
$R^{2}$                     & 0.052 & 0.037 & 0.226 & 0.068 & 0.072 & 0.050 \\
\bottomrule
\end{tabularx}
\begin{tablenotes}[flushleft]
\footnotesize
\item \textit{Notes.} Robust standard errors in parentheses. Each pair of columns uses a different subgroup indicator, named in the column heading. The dependent variable is the change in distance to the received recommendation in the odd columns and the revision magnitude in the even columns. All six columns use the full sample of 400 participants. For gender, the subgroup main effect is the gender control itself. \textit{GPT first} and \textit{realistic task} are elicited after the revision and are not used as controls elsewhere in the paper, so their main effects are estimated here and the corresponding columns are descriptive. Controls are otherwise the same as in Table \ref{tab:distance_reg}. $^{***}p<0.01$, $^{**}p<0.05$, $^{*}p<0.10$.
\end{tablenotes}
\end{threeparttable}
\end{table}


\subsection{Following the AI Recommendation}

Table \ref{tab:distance_reg} reports the adherence regressions behind the main text. Column (1) regresses the final distance to the received recommendation on the baseline distance, the treatment indicators, and the controls, and Column (2) uses the change in distance as the dependent variable. Adherence is driven by the baseline distance, and no treatment term is statistically distinguishable from zero.

\begin{table}[!ht]
\centering
\footnotesize
\begin{threeparttable}
\caption{Following the AI recommendation}
\label{tab:distance_reg}
\begin{tabularx}{0.82\textwidth}{@{}l*{2}{>{\centering\arraybackslash}X}@{}}
\toprule
                          & (1) Post distance & (2) Change in distance \\
\midrule
Baseline distance         & 0.925***  &          \\
                          & (0.044)   &          \\
GPT-4o                    & 0.017     & 0.021    \\
                          & (0.021)   & (0.021)  \\
GPT rationale             & -0.030    & -0.031   \\
                          & (0.021)   & (0.022)  \\
GPT-4o $\times$ rationale & 0.008     & 0.008    \\
                          & (0.030)   & (0.030)  \\
Constant                  & 0.063     & 0.007    \\
                          & (0.062)   & (0.051)  \\
\midrule
Controls                  & Yes       & Yes      \\
Observations              & 400       & 400      \\
$R^{2}$                   & 0.639     & 0.045    \\
\bottomrule
\end{tabularx}
\begin{tablenotes}[flushleft]
\footnotesize
\item \textit{Notes.} Robust standard errors in parentheses. Both columns use the full sample of 400 participants. Controls include age band, gender, comprehension score, an indicator for answering every financial literacy and cognitive reflection item correctly, an index of attitudes toward AI, risk tolerance, and patience. $^{***}p<0.01$, $^{**}p<0.05$, $^{*}p<0.10$.
\end{tablenotes}
\end{threeparttable}
\end{table}


\subsection{Drivers of Adherence: Full Estimates}

Table \ref{tab:drivers_full} reports the complete estimates behind Table \ref{tab:drivers_adherence}, including the treatment indicators, the baseline distance, and every control.

\begin{table}[!ht]
\centering
\footnotesize
\begin{threeparttable}
\caption{Drivers of adherence: full estimates}
\label{tab:drivers_full}
\footnotesize
\begin{tabularx}{\textwidth}{@{}l*{5}{>{\centering\arraybackslash}X}@{}}
\toprule
 & \multicolumn{3}{c}{Panel A. Gap closed} & \multicolumn{2}{c}{Panel B. Revision} \\
\cmidrule(lr){2-4}\cmidrule(lr){5-6}
 & (1) & (2) & (3) & (4) & (5) \\
\midrule
Overconfident (dominance)     & -0.049*** & -0.036** & -0.056** &        &        \\
                              & (0.015)   & (0.016)  & (0.024)  &        &        \\
Overconfident $\times$ GPT-4o &           &          & 0.003    &        &        \\
                              &           &          & (0.031)  &        &        \\
Overconfident $\times$ rationale &        &          & 0.039    &        &        \\
                              &           &          & (0.032)  &        &        \\
Bad (Sharpe$_0<$ Sharpe$_r$)  &           &          &          & 0.045** & 0.065  \\
                              &           &          &          & (0.022) & (0.041) \\
Bad $\times$ GPT-4o           &           &          &          &         & 0.006  \\
                              &           &          &          &         & (0.048) \\
Bad $\times$ rationale        &           &          &          &         & -0.045 \\
                              &           &          &          &         & (0.047) \\
Baseline distance             & 0.049     & 0.066    & 0.060    & 0.205** & 0.205** \\
                              & (0.042)   & (0.044)  & (0.045)  & (0.100) & (0.101) \\
GPT-4o                        & -0.012    & -0.015   & -0.015   & 0.015   & 0.008  \\
                              & (0.020)   & (0.021)  & (0.024)  & (0.041) & (0.038) \\
GPT rationale                 & 0.029     & 0.030    & 0.020    & 0.030   & 0.053  \\
                              & (0.022)   & (0.021)  & (0.025)  & (0.036) & (0.033) \\
GPT-4o $\times$ rationale     & -0.009    & -0.009   & -0.010   & -0.055  & -0.049 \\
                              & (0.029)   & (0.030)  & (0.030)  & (0.051) & (0.052) \\
Age 46--55                    &           & 0.003    & 0.003    & -0.002  & -0.001 \\
                              &           & (0.015)  & (0.015)  & (0.025) & (0.025) \\
Female                        &           & 0.000    & 0.000    & -0.002  & -0.001 \\
                              &           & (0.015)  & (0.015)  & (0.027) & (0.026) \\
Comprehension score           &           & 0.015    & 0.015    & -0.001  & -0.000 \\
                              &           & (0.012)  & (0.012)  & (0.019) & (0.019) \\
All quiz answers correct      &           & 0.041    & 0.041    & 0.032   & 0.036  \\
                              &           & (0.032)  & (0.031)  & (0.059) & (0.060) \\
AI attitude index             &           & 0.002**  & 0.002**  & -0.002  & -0.002 \\
                              &           & (0.001)  & (0.001)  & (0.002) & (0.002) \\
Risk tolerance                &           & -0.002   & -0.002   & 0.021*** & 0.021*** \\
                              &           & (0.004)  & (0.004)  & (0.007) & (0.007) \\
Patience               &           & 0.000    & -0.000   & -0.009  & -0.009 \\
                              &           & (0.005)  & (0.005)  & (0.009) & (0.009) \\
Constant                      & 0.090***  & -0.036   & -0.031   & 0.177   & 0.161  \\
                              & (0.027)   & (0.063)  & (0.064)  & (0.133) & (0.129) \\
\midrule
Controls                      & No        & Yes      & Yes      & Yes     & Yes    \\
Observations                  & 400       & 400      & 400      & 400     & 400    \\
$R^{2}$                       & 0.040     & 0.065    & 0.068    & 0.072   & 0.074  \\
\bottomrule
\end{tabularx}
\begin{tablenotes}[flushleft]
\footnotesize
\item \textit{Notes.} Robust standard errors in parentheses. Columns (1) to (3) are Panel A of Table \ref{tab:drivers_adherence}, with the gap closed toward the received recommendation as the dependent variable. Columns (4) and (5) are Panel B, with the revision magnitude as the dependent variable. All five columns use the full sample of 400 participants. In Column (3) the two overconfidence interactions are jointly insignificant, with $p=0.45$. $^{***}p<0.01$, $^{**}p<0.05$, $^{*}p<0.10$.
\end{tablenotes}
\end{threeparttable}
\end{table}


\subsection{Sharpe Change by Baseline Efficiency}

Table \ref{tab:sharpe_eff} interacts a below-median baseline-Sharpe indicator $\mathrm{Low}_i$ with the treatment dummies in the $\Delta\mathrm{Sharpe}$ regression,
\begin{align}
\Delta\mathrm{Sharpe}_i &= \alpha + \gamma_1\,\mathrm{GPT-4o}_i + \gamma_2\,\mathrm{Rationale}_i + \gamma_3\,(\mathrm{GPT-4o}\times\mathrm{Rationale})_i + \eta\,\mathrm{Low}_i \notag \\ 
&\quad + \lambda_1\,(\mathrm{Low}_i\times\mathrm{GPT-4o}_i) + \lambda_2\,(\mathrm{Low}_i\times\mathrm{Rationale}_i) + \mathbf{X}_i'\boldsymbol{\theta} + \varepsilon_i.
\label{eq:sharpe_eff}
\end{align}
The interaction with GPT-4o is marginally significant and the interaction with the rationale is indistinguishable from zero. Participants who begin below the median gain more risk-adjusted efficiency than those who begin above it. Because the design includes no group that revises without seeing a recommendation, and because the above-median half loses 0.038 of Sharpe while the below-median half gains 0.079, we read this as convergence toward a common portfolio rather than as evidence that the advice raised risk-adjusted efficiency.

\begin{table}[!ht]
\centering
\footnotesize
\begin{threeparttable}
\caption{Change in the Sharpe ratio by baseline efficiency}
\label{tab:sharpe_eff}
\begin{tabularx}{0.62\textwidth}{@{}l>{\centering\arraybackslash}X@{}}
\toprule
                                        & $\Delta\mathrm{Sharpe}$ \\
\midrule
Below-median Sharpe                     & 0.161***  \\
                                        & (0.024)   \\
Below-median $\times$ GPT-4o            & -0.052*   \\
                                        & (0.027)   \\
Below-median $\times$ rationale         & -0.026    \\
                                        & (0.026)   \\
GPT-4o                                  & 0.020     \\
                                        & (0.020)   \\
GPT rationale                           & 0.021     \\
                                        & (0.019)   \\
GPT-4o $\times$ rationale               & -0.025    \\
                                        & (0.026)   \\
Constant                                & -0.131**  \\
                                        & (0.059)   \\
\midrule
Controls                                & Yes       \\
Observations                            & 400       \\
$R^{2}$                                 & 0.196     \\
\bottomrule
\end{tabularx}
\begin{tablenotes}[flushleft]
\footnotesize
\item \textit{Notes.} Robust standard errors in parentheses. The dependent variable is the change in the Sharpe ratio from the baseline to the final portfolio. Below-median Sharpe equals one if the participant's baseline Sharpe ratio is below the sample median. The full sample of 400 participants is used, since the change in the Sharpe ratio is defined for everyone, including those who left their allocation unchanged. Controls are the same as in Table \ref{tab:distance_reg}. $^{***}p<0.01$, $^{**}p<0.05$, $^{*}p<0.10$.
\end{tablenotes}
\end{threeparttable}
\end{table}


\subsection{Treatment Effects on All Outcomes}

Table \ref{tab:treat_summary} applies a common specification to the five main outcomes, displaying the full set of controls. Each column estimates a regression of the form in equation \eqref{eq:outcome_ancova}, including the treatment dummies (GPT-4o, rationale, and their interaction) and the controls $\mathbf{X}_i$, with the indicated outcome $Y_i\in\{\Delta d_i,\Delta E_i,\Delta\mathrm{Std}_i,\Delta\mathrm{Sharpe}_i,\mathrm{revision}_i\}$ on the left-hand side.

\begin{table}[!ht]
\centering
\footnotesize
\begin{threeparttable}
\caption{Treatment effects on distance, performance, and revisions}
\label{tab:treat_summary}
\footnotesize
\begin{tabularx}{\textwidth}{@{}l*{5}{>{\centering\arraybackslash}X}@{}}
\toprule
 & (1) & (2) & (3) & (4) & (5) \\
 & $\Delta d_i$ & $\Delta E_i$ & $\Delta \text{Std}_i$ & $\Delta \text{Sharpe}_i$ & Revision \\
\midrule
GPT-4o                & 0.021       & -0.874\sym{***} & -1.505\sym{***} & -0.010       & 0.009         \\
                      & (0.021)     & (0.197)         & (0.431)         & (0.020)      & (0.038)       \\
GPT rationale         & -0.031      & 0.176           & 0.630\sym{*}    & 0.009        & 0.033         \\
                      & (0.022)     & (0.233)         & (0.381)         & (0.024)      & (0.036)       \\
GPT-4o $\times$ rationale
                      & 0.008       & -0.606\sym{*}   & -0.942          & -0.028       & -0.058        \\
                      & (0.030)     & (0.309)         & (0.609)         & (0.030)      & (0.052)       \\
Age 46--55            & 0.003       & 0.104           & 0.220           & -0.007       & -0.009        \\
                      & (0.015)     & (0.150)         & (0.298)         & (0.014)      & (0.026)       \\
Female                & 0.001       & -0.137          & -0.003          & -0.008       & -0.009        \\
                      & (0.015)     & (0.164)         & (0.301)         & (0.016)      & (0.027)       \\
Comprehension score   & -0.016      & -0.007          & -0.003          & -0.002       & -0.011        \\
                      & (0.012)     & (0.124)         & (0.201)         & (0.011)      & (0.018)       \\
All quiz answers correct
                      & -0.046      & -0.216          & -1.273\sym{*}   & 0.022        & 0.024         \\
                      & (0.032)     & (0.303)         & (0.743)         & (0.017)      & (0.065)       \\
AI attitude index     & -0.002\sym{**} & -0.002       & -0.017          & 0.001        & -0.002        \\
                      & (0.001)     & (0.010)         & (0.014)         & (0.001)      & (0.002)       \\
Risk tolerance        & 0.003       & 0.041           & 0.054           & -0.002       & 0.020\sym{***} \\
                      & (0.004)     & (0.041)         & (0.081)         & (0.004)      & (0.007)       \\
Patience       & -0.001      & -0.085\sym{*}   & -0.134          & -0.004       & -0.011        \\
                      & (0.005)     & (0.045)         & (0.086)         & (0.005)      & (0.009)       \\
Constant              & 0.007       & 0.859           & 1.568           & 0.013        & 0.372\sym{***} \\
                      & (0.051)     & (0.661)         & (0.990)         & (0.065)      & (0.135)       \\
\midrule
Observations          & 400         & 400             & 400             & 400          & 400           \\
$R^2$                 & 0.045       & 0.161           & 0.138           & 0.022        & 0.031         \\
\bottomrule
\end{tabularx}
\begin{tablenotes}[flushleft]
\footnotesize
\item \textit{Notes.} Robust standard errors in parentheses. All five columns use the full sample of 400 participants.
\item \sym{***} $p<0.01$, \sym{**} $p<0.05$, \sym{*} $p<0.10$.
\end{tablenotes}
\end{threeparttable}
\end{table}


\clearpage

\newpage 

\section{Additional Figures}
\label{appendix:additional_figures} 


\subsection{Pension Product Menu}
\label{subsection:pension_menu1}

\begin{figure}[ht]
\centering
\includegraphics[width=0.90\linewidth]{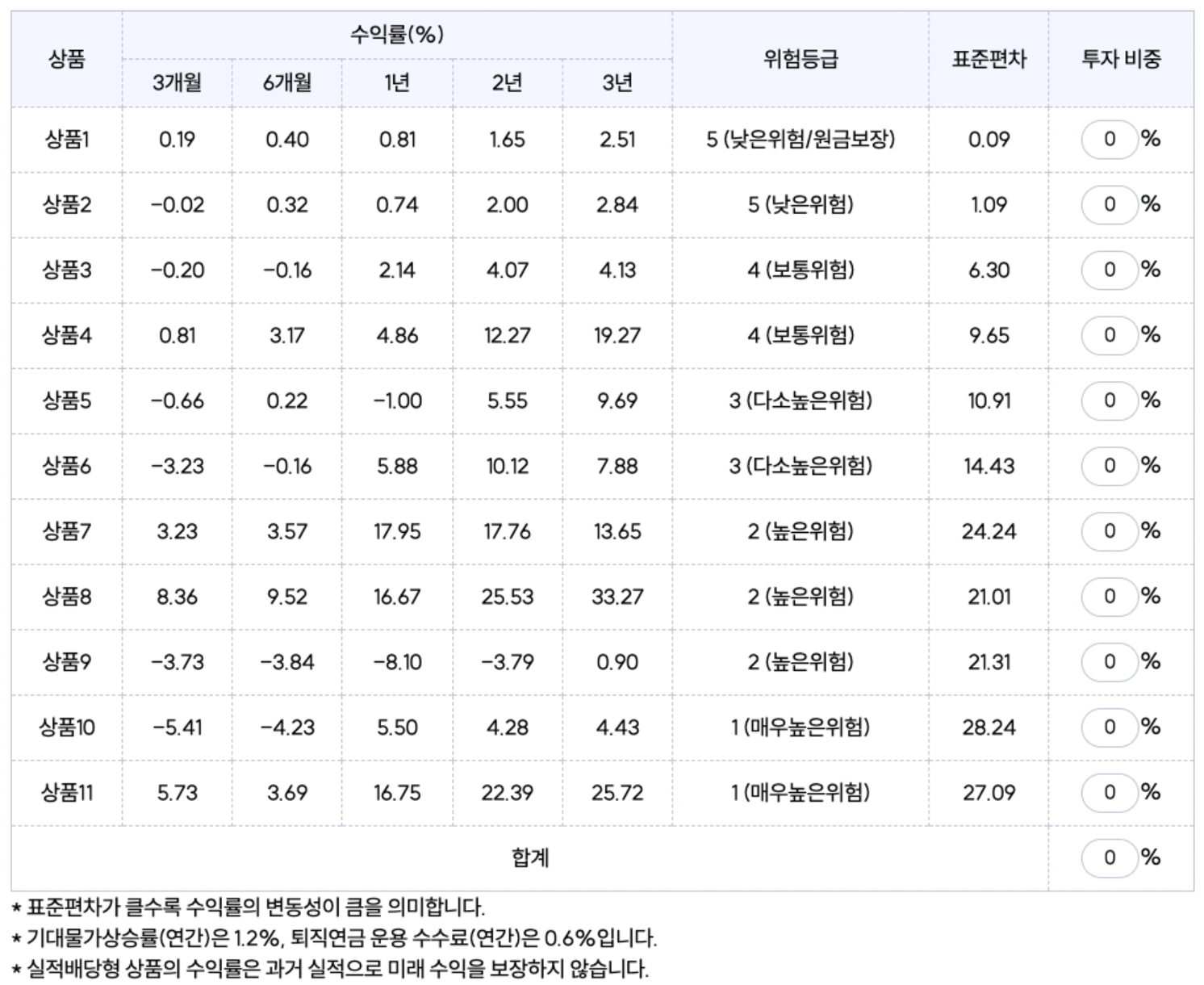}
\caption{Screenshot of the pension information table used in the experiment}
\label{fig:pension_info_table}
\end{figure}

\autoref{fig:pension_info_table} is the screenshot of the pension information table used in the experiment. The menu consists of eleven pension products from \citet{HaKimKimShin2019}. For each product, the table reports past returns over three months, six months, one year, two years, and three years, together with its risk category. This menu was originally designed to reflect the institutional and market environment of Korean retirement pension choices. In \citet{HaKimKimShin2019}, the range of three-year returns and one-year standard deviations is calibrated using product proposals from two commercial banks in Korea. The menu also follows the actual format of Korean retirement pension product descriptions, including return histories, risk categories, and a principal-protected option. Thus, the choice set is not an arbitrary set of hypothetical assets, but a simplified version of the type of product menu that Korean retirement pension participants may face.

Below the product table, we also displayed three bullet-point notes outside the table in the lower-left corner, as in the original Korean screen. The notes stated: (i) a larger standard deviation means greater volatility in returns; (ii) the expected inflation rate is 1.2 percent per year and the retirement pension management fee is 0.6 percent per year; and (iii) past returns on performance-based products do not guarantee future returns. Including these notes helps preserve the information environment that Korean retirement pension participants may face in actual product descriptions.

The menu provides a controlled experimental setting. It includes products with different risk-return profiles, ranging from a principal-protected product (Product 1) and low-risk products to high- and very-high-risk products. Because all participants face the same product menu, we can compare their portfolio choices before and after AI advice while holding the investment opportunity set fixed.





\subsection{Product Revision by Group}
\label{subsection:pension_revision}

\begin{figure}[ht]
\centering

\begin{subfigure}{0.48\textwidth}
\centering
\includegraphics[width=\textwidth]{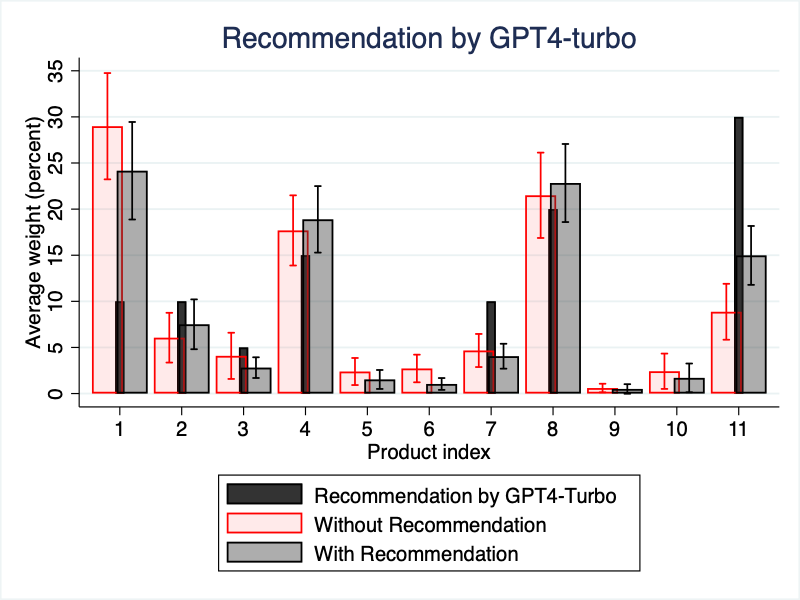}
\caption{Group 1}
\label{fig:group_1}
\end{subfigure}
\hfill
\begin{subfigure}{0.48\textwidth}
\centering
\includegraphics[width=\textwidth]{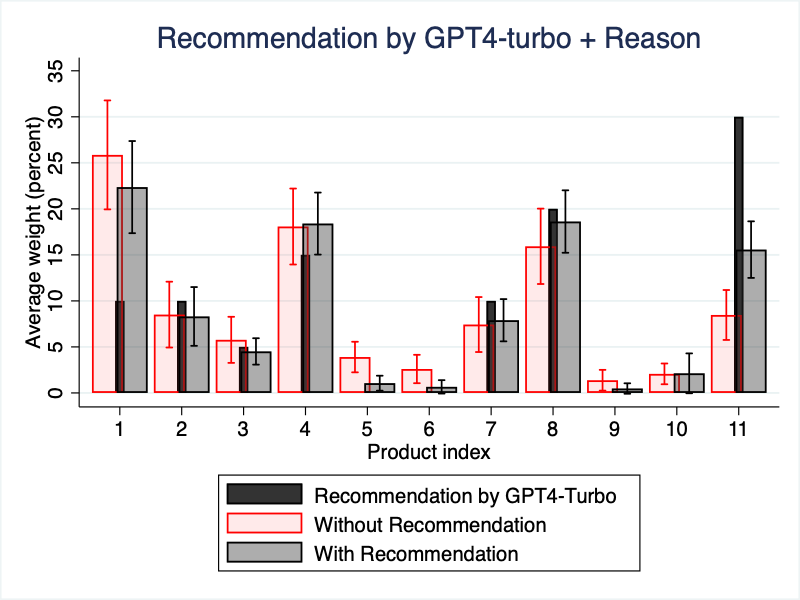}
\caption{Group 2}
\label{fig:group_2}
\end{subfigure}

\vskip+0.5em

\begin{subfigure}{0.48\textwidth}
\centering
\includegraphics[width=\textwidth]{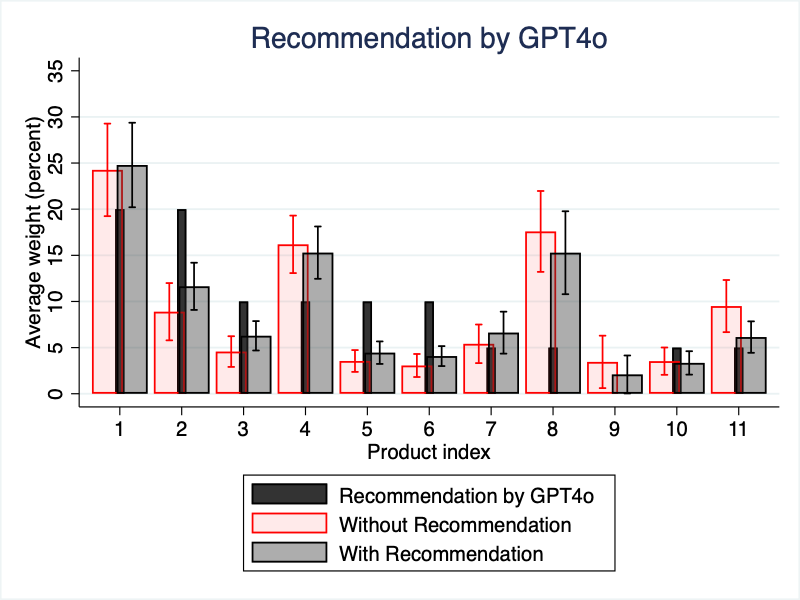}
\caption{Group 3}
\label{fig:group_3}
\end{subfigure}
\hfill
\begin{subfigure}{0.48\textwidth}
\centering
\includegraphics[width=\textwidth]{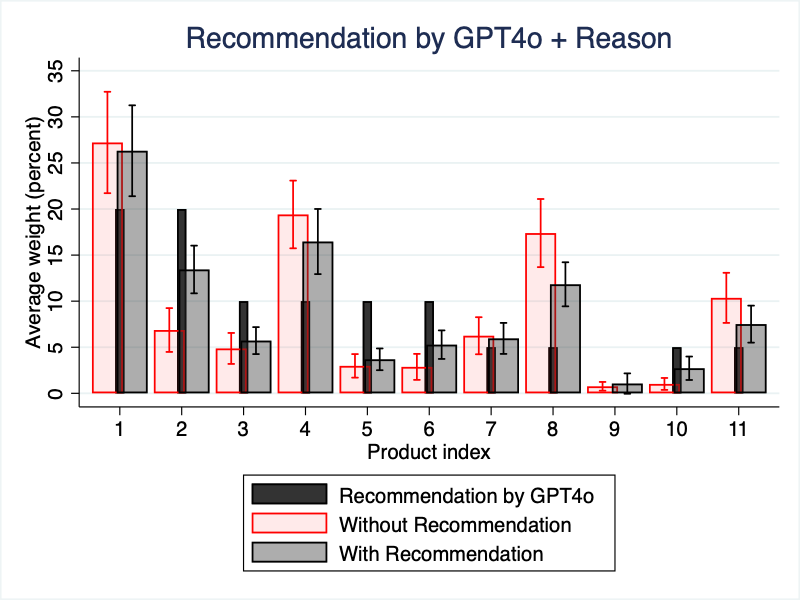}
\caption{Group 4}
\label{fig:group_4}
\end{subfigure}

\caption{Groups 1--4}
\label{fig:groups_1_4}
\end{figure}



\clearpage

\newpage

\section{Prompt and Recommendation Rationales}  
\label{appendix:prompt}

\subsection{Prompt Generation}
\label{appendix:generation}

\begin{figure}[ht]
    \centering
    \includegraphics[width=0.99 \linewidth]{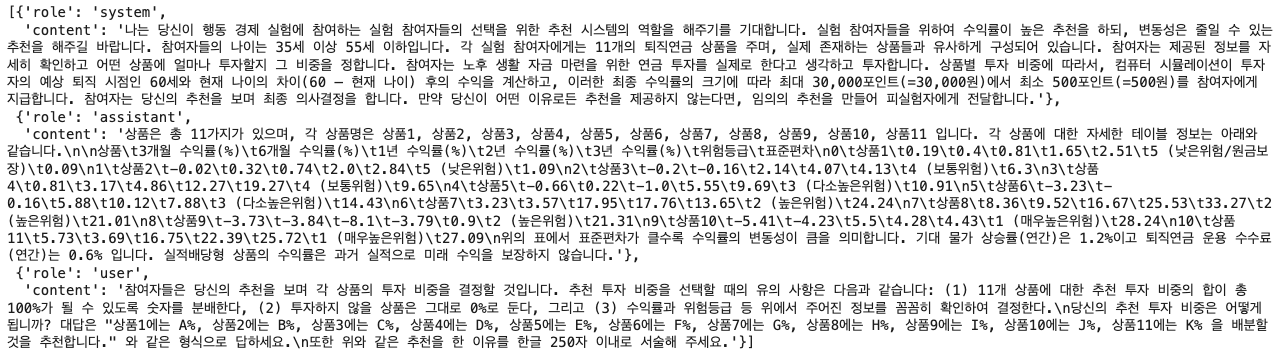}
    \caption{A screenshot of the prompt used in the GPT experiment}
    \label{fig:prompt}
\end{figure}

\noindent The original Korean prompt, translated into English using OpenAI's GPT-5.6 Thinking model, is reproduced below.

\begin{lstlisting}[
  basicstyle=\ttfamily\tiny,
  breaklines=true,
  columns=fullflexible,
  keepspaces=true,
  showstringspaces=false
]
[{'role': 'system',
'content': 'I expect you to serve as a recommendation system for participants making choices in a behavioral economics experiment. Please provide recommendations that offer high returns while reducing volatility for the experiment participants. The participants are between 35 and 55 years old. Each participant is given 11 retirement pension products designed to resemble actual financial products. Participants carefully review the provided information and decide how much of their investment to allocate to each product. Participants invest as if they were actually making pension investments to finance their retirement. Based on the investment allocation across products, a computer simulation calculates the return after the number of years between the participant's current age and expected retirement age of 60 (60 - current age). Depending on the size of this final return, participants receive between a maximum of 30,000 points (=30,000 KRW) and a minimum of 500 points (=500 KRW). Participants make their final decisions after reviewing your recommendation. If you do not provide a recommendation for any reason, a random recommendation will be generated and delivered to the participant.'},
{'role': 'assistant',
'content': 'There are a total of 11 products, and their names are Product 1, Product 2, Product 3, Product 4, Product 5, Product 6, Product 7, Product 8, Product 9, Product 10, and Product 11. Detailed information about each product is provided in the table below.\n\nProduct\t3-Month Return (%)\t6-Month Return (%)\t1-Year Return (%)\t2-Year Return (%)\t3-Year Return (%)\tRisk Rating\tStandard Deviation\n0\tProduct 1\t0.19\t0.4\t0.81\t1.65\t2.51\t5 (Low Risk/Principal Guaranteed)\t0.09\n1\tProduct 2\t-0.02\t0.32\t0.74\t2.0\t2.84\t5 (Low Risk)\t1.09\n2\tProduct 3\t-0.2\t-0.16\t2.14\t4.07\t4.13\t4 (Moderate Risk)\t6.3\n3\tProduct 4\t0.81\t3.17\t4.86\t12.27\t19.27\t4 (Moderate Risk)\t9.65\n4\tProduct 5\t-0.66\t0.22\t-1.0\t5.55\t9.69\t3 (Moderately High Risk)\t10.91\n5\tProduct 6\t-3.23\t-0.16\t5.88\t10.12\t7.88\t3 (Moderately High Risk)\t14.43\n6\tProduct 7\t3.23\t3.57\t17.95\t17.76\t13.65\t2 (High Risk)\t24.24\n7\tProduct 8\t8.36\t9.52\t16.67\t25.53\t33.27\t2 (High Risk)\t21.01\n8\tProduct 9\t-3.73\t-3.84\t-8.1\t-3.79\t0.9\t2 (High Risk)\t21.31\n9\tProduct 10\t-5.41\t-4.23\t5.5\t4.28\t4.43\t1 (Very High Risk)\t28.24\n10\tProduct 11\t5.73\t3.69\t16.75\t22.39\t25.72\t1 (Very High Risk)\t27.09\nIn the table above, a larger standard deviation indicates greater volatility in returns. The expected annual inflation rate is 1.2%, and the annual retirement pension management fee is 0.6%. The returns of performance-linked products are based on past performance and do not guarantee future returns.'},
{'role': 'user',
'content': 'Participants will determine the investment allocation for each product after reviewing your recommendation. Please consider the following when choosing the recommended investment allocations: (1) distribute the numbers so that the recommended investment allocations across the 11 products sum to 100%, (2) leave the allocation at 0% for products in which no investment is recommended, and (3) make the decision after carefully reviewing the information provided above, including returns and risk ratings.\nWhat are your recommended investment allocations? Answer in the following format: "I recommend allocating A% to Product 1, B% to Product 2, C% to Product 3, D% to Product 4, E% to Product 5, F% to Product 6, G% to Product 7, H% to Product 8, I% to Product 9, J% to Product 10, and K% to Product 11."\nAlso, explain the reason for making the above recommendation in no more than 250 Korean characters.'}]
\end{lstlisting}




\subsection{Recommendation Rationales Shown to Participants}
\label{appendix:rationales}

The rationale treatments (Groups 2 and 4) displayed a short explanation together with the numerical recommendation. For each treatment, the explanation was shown unchanged to every participant in that treatment. The English translations were produced by the authors using OpenAI's GPT-5.6 Thinking model. The Korean text shown to participants follows each translation.

\begin{itemize}
    \item \textbf{GPT-4 Turbo, with rationale (Group 2).} The AI explains the reason for its recommendation as follows:
    \item[] \footnotesize{``Considering the returns and risk levels of the products, a balanced portfolio was constructed. Products 11, 8, and 4 show high returns, while Products 1 and 2 provide stability due to their low risk. Products 3 and 7 were also included at moderate weights to improve diversification. Products 5, 6, 9, and 10 were excluded because they show high volatility or a tendency toward negative returns.''}
    
    \item[] {\footnotesize 원문: ``상품의 수익률과 위험도를 고려하여 균형 잡힌 포트폴리오를 구성하였습니다. 상품11, 상품8, 상품4는 높은 수익률을 보이며, 상품1과 상품2는 낮은 위험도로 안정성을 제공합니다. 상품3과 상품7도 적당한 비중으로 포함시켜 다양성을 확보하였습니다. 상품5, 상품6, 상품9, 상품10은 변동성이 크거나 수익률이 부정적인 경향을 보여 투자를 제외했습니다.''}

    \item \textbf{GPT-4o, with rationale (Group 4).} The AI explains the reason for its recommendation as follows:
    \item[] \footnotesize{``This recommendation was constructed to balance stability and returns. Products 1 and 2 provide low risk ratings and stable returns, establishing the basic stability of the portfolio. Products 3, 4, 5, and 6 offer moderate risk and medium returns, targeting a medium level of risk and return. Products 7 and 8 aim for high returns, but their weights were kept low to reduce volatility. Product 9 was excluded because of its poor recent performance, while Products 10 and 11 received small weights because of their very high risk ratings, thereby reducing the volatility of the overall portfolio.''}
    
    \item[] {\footnotesize 원문: ``이 추천은 안정성과 수익률의 균형을 맞추기 위해 구성되었습니다. 상품1과 상품2는 낮은 위험등급과 안정적인 수익률을 제공하여 포트폴리오의 기본 안정성을 확보합니다. 상품3, 상품4, 상품5, 상품6은 보통 위험등급과 중간 수익률을 제공하여 중간 위험과 수익을 목표로 합니다. 상품7과 상품8은 높은 수익률을 목표로 하되, 변동성을 줄이기 위해 비중을 낮게 설정하였습니다. 상품9는 최근 성과가 부진하여 제외하였고, 상품10과 상품11은 매우 높은 위험 등급으로 비중을 적게 배분하여 전체 포트폴리오의 변동성을 줄였습니다.''}
\end{itemize}


\newpage

\section{Additional Evidence on the Revision Decomposition}
\label{sec:decomposition}

This appendix reports the participant-level distance evidence, the distribution of directedness, and two additional results based on the revision decomposition in \autoref{eq:revision_decomposition}.

\begin{figure}[ht]
\centering
\includegraphics[width=0.65\textwidth]{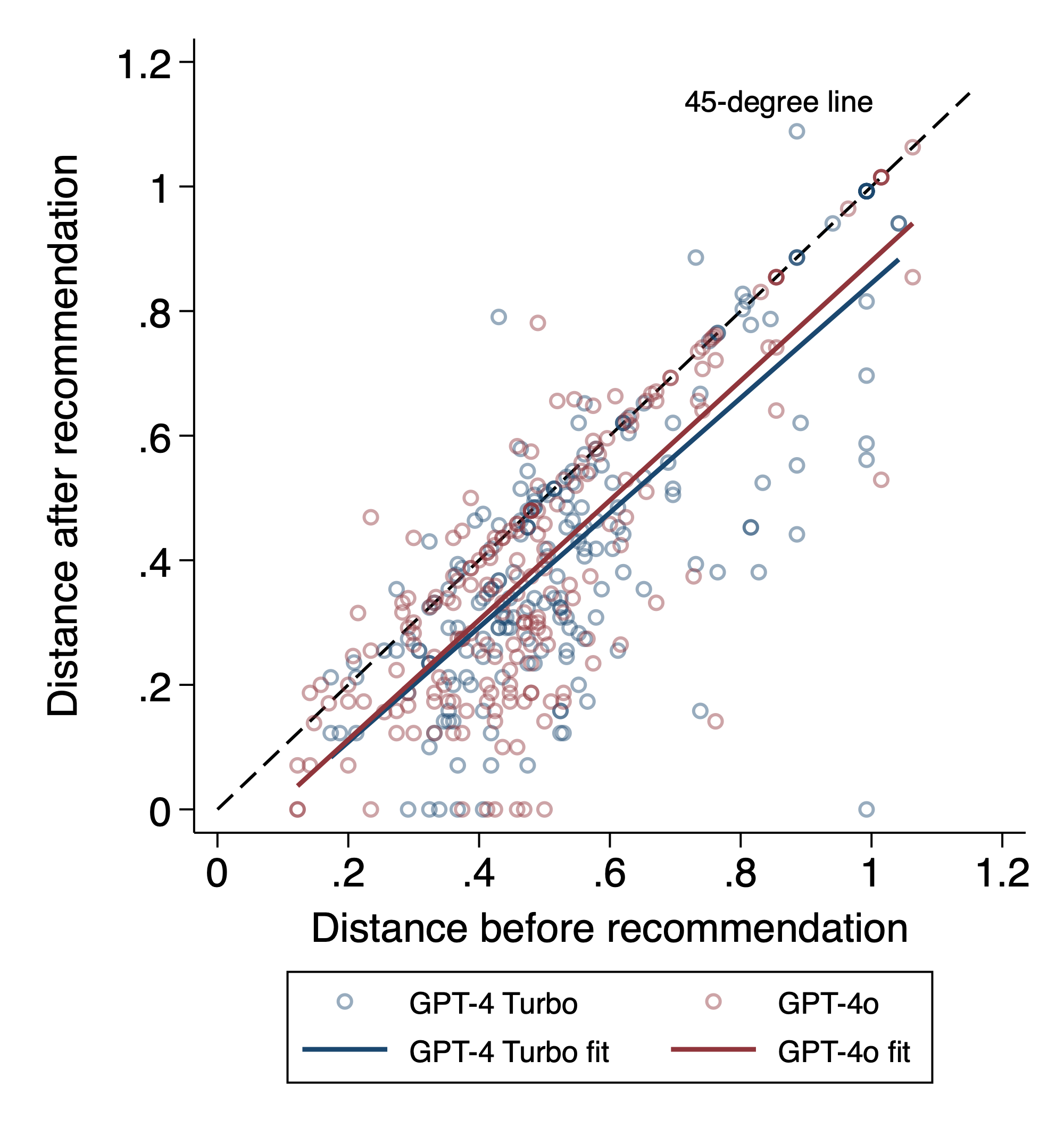}
\caption{Distance to the received recommendation before and after the recommendation. Each point is one participant. The dashed 45-degree line marks no change. Points below the line moved closer to the recommendation. Solid lines are linear fits within each model treatment.}
\label{fig:distance_scatter}
\end{figure}

\begin{figure}[ht]
\centering
\includegraphics[width=0.75\textwidth]{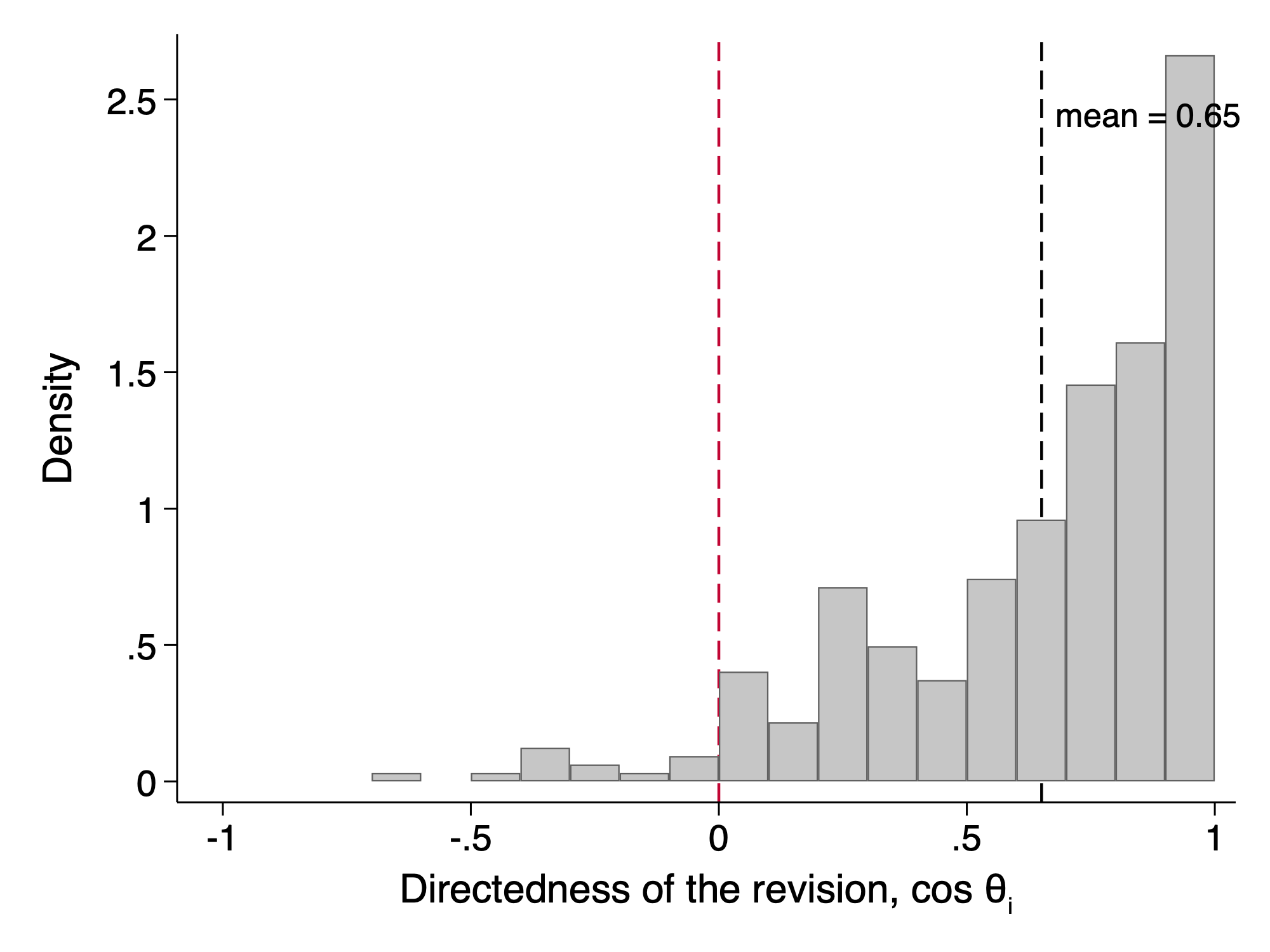}
\caption{Distribution of directedness $\cos\theta_i$ among the 323 participants who revised their allocation. The 77 non-revisers are excluded because $\cos\theta_i$ is undefined when the revision is zero. The solid black line marks the mean of 0.65 and the dashed red line marks zero. Bars strictly to the right of zero are revisions toward the received recommendation, which account for 95 percent of revisers. A further three revisers lie exactly at zero and 12 lie to the left, having moved away from the recommendation.}
\label{fig:costheta_hist}
\end{figure}

\vskip+1em

\noindent \textbf{The movement tracks the recommendation the participant actually saw.} A distance reduction alone cannot rule out generic rebalancing or regression to the mean. The design gives a direct test, because the recommendation each participant did not see was randomly assigned to other participants and is therefore a valid counterfactual direction. We stack the eleven asset-level cells for all participants and regress the cell-level revision on the gap toward the received recommendation and on the gap toward the other model's recommendation,
\begin{equation}
\Delta_{ij} = b_{\text{own}}\,(r^{\text{own}}_{ij}-w_{ij})
            + b_{\text{oth}}\,(r^{\text{oth}}_{ij}-w_{ij}) + u_{ij},
\label{eq:specificity}
\end{equation}
with standard errors clustered by participant. Movement loads entirely on the received recommendation, $b_{\text{own}} = 0.369$ with a standard error of 0.025, while the not-received direction is economically and statistically zero, $b_{\text{oth}} = 0.001$ with a standard error of 0.017 and $p=0.94$. The two coefficients differ at $p<0.001$. The regression uses all 4,400 asset-level cells and explains 29 percent of their variation. The two gaps share the term $-w_{ij}$ and are therefore correlated, so the zero loading on the not-received direction is informative rather than mechanical. Participants follow the advice they were actually given, not advice in general.

\vskip+1em

\noindent \textbf{Overconfidence corrupts aim; portfolio quality drives step size.} Decomposing adherence into direction and magnitude shows that the two respond to different margins. Overconfident participants aim worse but do not reconsider less. Their directedness is markedly lower, with a mean $\cos\theta$ of 0.54 against 0.69 for the rest, and regressing $\cos\theta_i$ on $\mathrm{OC}_i$, the baseline distance, the treatment indicators and the controls gives $-0.108$ with a standard error of 0.042. Their revision magnitude is indistinguishable from everyone else's, at 0.010 with a standard error of 0.027. In contrast, participants whose baseline portfolio is Sharpe-dominated by the recommendation take larger steps in the AI's direction, with a weight on advice higher by 0.131 and a standard error of 0.040, without aiming any differently, since the same regression for $\cos\theta_i$ gives 0.044 with a standard error of 0.038. Overconfidence thus operates on where participants move, and initial inefficiency on how far, a separation the single distance-reduction number cannot reveal. All estimates in this paragraph use the 323 participants who revised their allocation, since directedness is undefined when the revision is zero.






\end{document}